\documentstyle[psfig]{mn}

\def\etal{{\it et al.\ }}
\def\eg{{\it e.g.\ }}

\def\ie{{\it i.e.\ }}

\def\spose#1{\hbox to 0pt{#1\hss}}
\def\approxlt{\mathrel{\spose{\lower 3pt\hbox{$\sim$}}
	\raise 2.0pt\hbox{$<$}}}
\def\approxgt{\mathrel{\spose{\lower 3pt\hbox{$\sim$}}
	\raise 2.0pt\hbox{$>$}}}
\def\approxpropto{\mathrel{\spose{\lower 3pt\hbox{$\sim$}}
	\raise 2.0pt\hbox{$\propto$}}}
\mathchardef\twiddle="2218

\def\multleft#1{\hbox to size{\vbox {\halign {\lft{##}\cr #1}}\hfill}\par}
\def\multright#1{\hbox to size{\vbox {\halign {\rt{##}\cr #1}}\hfill}\par}

\def\today{\ifcase\month\or January\or February\or March\or April\or May\or
      June\or July\or August\or September\or October\or November\or December\fi
      \space\number\day, \number\year}
\def\<{\thinspace}

\def\apc{\rm atom cm$^{-2}$}

\def\erg{{\rm\thinspace erg}}

\def\keV{{\rm\thinspace keV}}

\def\km{{\rm\thinspace km}}

\def\Mpc{{\rm\thinspace Mpc}}
\def\Msun{\hbox{$\rm\thinspace M_{\odot}$}}

\def\s{{\rm\thinspace s}}
\def\yr{{\rm\thinspace yr}}

\def\ergps{\hbox{$\erg\s^{-1}\,$}}

\def\kmps{\hbox{$\km\s^{-1}\,$}}

\def\Msunpyr{\hbox{$\Msun\yr^{-1}\,$}}

\def\kmpspMpc{\hbox{$\kmps\Mpc^{-1}$}}

\def\apc{\rm atom cm$^{-2}$}
\title[The properties of cooling flows in X-ray luminous clusters of
galaxies]
{The properties of cooling flows in X-ray luminous clusters of galaxies}
\author[S.W. Allen ]
{\parbox[]{6.in} {S.W. Allen \\
\footnotesize
Institute of Astronomy, Madingley Road, Cambridge CB3 OHA\\
}}
\begin{document}
\maketitle
\begin{abstract}
We discuss the X-ray properties of the cooling flows in a sample of thirty 
highly X-ray luminous clusters of galaxies observed with the ASCA and ROSAT
satellites. We demonstrate the need for multiphase models to consistently 
explain the spectral and imaging X-ray data for the clusters. The mass 
deposition rates of the cooling flows, independently determined from 
the ASCA spectra and ROSAT images, exhibit good agreement and exceed 1000 
\Msunpyr~in the largest systems. We confirm the presence of 
intrinsic X-ray absorption in the clusters using a variety of 
spectral models. The measured equivalent hydrogen 
column densities of absorbing material are 
sensitive to the spectral models used in the analysis, varying from average 
values of a few $10^{20}$ \apc~for a simple isothermal emission model, to a 
few $10^{21}$ \apc~using our preferred cooling flow models, assuming in each
case that the absorber lies in a uniform foreground screen. The true intrinsic 
column densities are likely to be even higher if the absorbing medium is
distributed throughout the clusters. We summarize the constraints on the 
nature of the X-ray absorber from observations in other wavebands. A 
significant part of the X-ray absorption may be due to dust. 

\end{abstract}


\section{Introduction}

X-ray observations of clusters of galaxies show that in the central
regions of most clusters the cooling time of the intracluster gas
is significantly less than a Hubble time (\eg Edge, Stewart \& Fabian 1992; 
White, Jones \& Forman 1997; Peres \etal 1998).  This cooling is thought to 
lead to a slow net inflow of material towards the cluster centre; a 
process known as a cooling flow (see Fabian 1994 for a review). 
X-ray imaging data show that gas typically
`cools out' throughout the central few tens to hundreds of kpc in 
clusters, with ${\dot M}(r) \approxpropto r$, where ${\dot M}(r)$ is the 
integrated mass deposition rate within radius $r$ (\eg Thomas, Fabian \&
Nulsen 1987). Spatially resolved X-ray spectroscopy has confirmed the 
presence of distributed cooling gas in cooling flows, with a 
spatial distribution and luminosity in good agreement with the 
predictions from the imaging data and cooling-flow models (\eg 
Allen \& Fabian 1997). 

For some years, the primary uncertainty with the standard model for 
cooling flows was the fate of the cooled matter
(\eg see Fabian, Nulsen \& Canizares 1991). However, the discovery of 
large column densities of intrinsic X-ray absorbing 
material associated with cooling flows observed with 
the Solid State Spectrometer (SSS) on the Einstein Observatory (White \etal
1991; Johnstone \etal 1992) opened one interesting possibility.
Follow-up spatially-resolved X-ray spectroscopy with the Position
Sensitive Proportional Counter (PSPC) on ROSAT (\eg Allen \etal 1993; Irwin 
\& Sarazin 1995; Allen \& Fabian 1997) confirmed the presence of intrinsic 
X-ray absorbing material in cooling-flow clusters and further showed this 
material to be distributed throughout, but centrally-concentrated 
within, the cooling flows. The X-ray data thus 
identify  the intrinsic X-ray absorbing material as a plausible sink for
the cooled gas deposited by the cooling flows.

In this paper we examine the X-ray properties of the cooling flows in a sample
of 30 of the most X-ray luminous ($L_{\rm Bol} > 10^{45}$ \ergps) 
clusters of galaxies known. Using ASCA spectra and ROSAT High
Resolution Imager (HRI) data, we present independent determinations
of the mass deposition rates in the cooling flows and measure the 
column densities of intrinsic X-ray absorbing material associated with 
these systems. We show that the cooling flow model provides a consistent
description of the X-ray imaging and spectral data and
that the observed masses of intrinsic X-ray absorbing
material are in reasonable agreement with the masses expected to have been 
accumulated by the cooling flows over their lifetimes. We also discuss 
the evidence for mass deposition from the cooling flows in other
wavebands. 

This is the final paper in a 
series that has examined the impact of cooling flows on 
mass measurements (Allen 1998), the $kT_{\rm X}-L_{\rm Bol}$ relation 
(Allen \& Fabian 1998a) and metallicity measurements (Allen \& Fabian 1998b) 
for X-ray luminous clusters. In this paper, we describe  
the data reduction and analysis procedures used in these works. 
A large number of other, previous studies have also 
analysed one or more of data sets included here
(see references in the papers listed above), although these studies have 
not, in general, examined the properties of the cooling flows in
the clusters. Exceptions are the deprojection analysis of Peres 
\etal (1998), and the combined ASCA/ROSAT studies of Allen \etal (1996), 
Schindler \etal (1997), B\"ohringer \etal (1998) and Rizza \etal (1998). 

The cosmological parameters $H_0$=50 \kmpspMpc, $\Omega = 1$ 
and $\Lambda = 0$ are assumed throughout.

\section{Observations and data reduction}

\subsection{Sample selection}

Our sample was identified from clusters contained in the 
ROSAT X-ray Brightest Abell-type 
Cluster Sample (XBACS; Ebeling \etal 1996) and 
ROSAT Brightest Cluster Sample (BCS; Ebeling \etal 1998) with 
X-ray luminosities in the $0.1-2.4$ keV band exceeding $10^{45}$ 
\ergps. From this initial list, we selected for study 
those targets with ASCA X-ray spectra and ROSAT High Resolution Imager 
(HRI) data available on the Goddard Space Flight  Centre (GSFC) public 
archive as of 1997 July 15. We supplemented this sample with a number of 
other, southern X-ray luminous clusters, known to exhibit strong gravitational 
lensing effects (PKS0745-191, RXJ1347.5-1145, MS2137.3-2353, AC114) and the 
distant, luminous cooling-flow clusters Abell 1068, 1704 and IRAS 09104+4109.  

The primary goal of this project was to investigate the 
X-ray properties of cooling flows in X-ray luminous clusters 
and their effects on the integrated 
X-ray properties of their host systems. We have not included either 
the Perseus (Abell 426) or Coma (Abell 1656) clusters in our study, 
although these systems met our selection criteria, since both 
are too close ($z=0.0183$ and $0.0232$, respectively) to allow their 
integrated cluster properties to be studied using the techniques 
used in this paper. Detailed analyses of these clusters and other 
nearby cooling flows are discussed by Allen \etal (1999).

The clusters Abell 370, 115 and 1758 also met the selection 
criteria but were not included in our study since 
ROSAT HRI images show them not be single, coherent 
structures but to consist of a number of merging subunits. 
The integrated X-ray spectra for such clusters 
will relate to the virial properties of the individual subclusters rather than 
the systems as a whole, and the assumptions of spherical symmetry and 
hydrostatic equilibrium required by the X-ray modeling will not apply.  
The X-ray images for the other clusters included in the sample do not exhibit 
any dramatic substructure that would suggest these assumptions to be
invalid. Our final sample consists of 30 clusters, spanning the redshift range
$0.056 < z < 0.451$, with a mean redshift of 0.21.

\subsection{The ASCA observations }

The ASCA (Tanaka, Inoue \& Holt 1994) observations were made over a 
three-and-a-half year period between 1993 April and 1996 December. 
The ASCA X-ray Telescope array (XRT) consists of four 
nested-foil telescopes, each focussed onto one of four detectors; two X-ray 
CCD cameras, the Solid-state Imaging Spectrometers (SIS; S0 and S1), and 
two Gas scintillation Imaging  Spectrometers (GIS; G2 and G3). The XRT 
provides a spatial resolution of $\sim 3$ arcmin Half Power Diameter 
(HPD) in the energy range $0.3 - 12$ keV.  The SIS detectors provide good 
spectral resolution [$\Delta E/E = 0.02(E/5.9 {\rm keV})^{-0.5}$] over a 
$22 \times 22$ arcmin$^2$ field of view. The GIS detectors provide poorer 
energy resolution [$\Delta E/E = 0.08(E/5.9 {\rm keV})^{-0.5}$] but cover
a larger circular field of view of $\sim 50$ arcmin diameter. 

For our analysis we have used the screened event lists from the
rev1 processing of the data sets available on the GSFC 
ASCA archive (for a detailed description of the rev1 processing 
see the GSFC ASCA Data Reduction Guide, published by GSFC.) 
The ASCA data were reduced using the FTOOLS software (version 3.6) issued 
by GSFC, from within the XSELECT environment (version 1.3). 
Further data-cleaning procedures as recommended in the ASCA Data Reduction 
Guide, including appropriate grade selection, gain  corrections and  manual
screening based on the individual instrument light curves, were followed. 
A summary of the ASCA observations, including the individual
exposure times after all screening procedures were carried
out, is provided in Table 1. 

Spectra were extracted from all four ASCA detectors (except 
in those few cases where the S1 data were lost due 
to saturation problems caused by flickering pixels in the
CCDs). The spectra were extracted from circular regions, centred on 
the peaks of the X-ray emission. For the SIS data, the radii of the regions 
used were selected to minimize the number of chip boundaries crossed 
(thereby minimizing the systematic uncertainties introduced by such crossings) 
whilst covering as large a region of the clusters as possible. 
Data from the regions between the chips were masked out and 
excluded. The final extraction radii for the SIS data are summarized in Table
2. Also included in that Table are the chip modes used in the observations 
(whether 1,2 or 4 chip mode) and the number of chips from which the 
extracted data were drawn. For the GIS data a constant extraction radius of 6 
arcmin was used. [We note that for Abell 2142, the 2 arcmin (3 arcmin) 
radius region surrounding the X-ray bright Seyfert-1 galaxy 1556+274 (offset 
by $\sim 4$ arcmin from the X-ray centre of the cluster) was masked out 
and excluded from the analysis of the SIS (GIS) data.]

For the GIS observations, and SIS observations of clusters in 
regions of low Galactic column density
($N_{\rm H} \approxlt 5 \times 10^{20}$ \apc), background subtraction was 
carried out using the `blank sky' observations 
of high Galactic latitude fields compiled during the 
performance verification stage of the ASCA
mission. For such data sets, the blank-sky observations provide a reasonable
representation of the cosmic and instrumental backgrounds in the
detectors. The background data were screened and grade selected in the same 
manner as the target observations and background 
spectra were extracted from the same regions of the detectors as the 
cluster spectra.  For the SIS observations of clusters in directions of 
higher Galactic column density, background spectra were 
extracted from regions of the chips that were relatively free from 
foreground cluster emission.

For the SIS data, response matrices were generated using the FTOOLS
SISRMG software. Where the spectra covered more than one chip, response 
matrices were created for each chip, which were then combined to form a
counts-weighted mean matrix. For the GIS analysis, the  response matrices 
issued by GSFC on 1995 March 6 were used.

\subsection{The ROSAT observations}

The ROSAT HRI observations were carried out between 1991 November and 1995
June. The HRI provides a $\sim 5$ arcsec (FWHM) X-ray imaging 
facility covering a $\sim 40 \times 40$ arcmin$^2$ field of view (David
\etal 1996). Reduction of the data was carried out with the Starlink 
ASTERIX package. X-ray images were created on a $2 \times 2$ arcsec$^2$ 
pixel scale, from which centres for the cluster X-ray emission 
were determined. Where more than one observation of a source was made, a
mosaic was constructed from the individual observations. 
For the cooling flow and intermediate clusters, the X-ray centres were 
identified from the peaks of the X-ray surface brightness distributions 
(which are easily determined from the HRI images). 
For the non-cooling flow clusters, the X-ray emission is not as sharply-peaked
and for these systems we have identified the X-ray centres with the 
results from iterative determinations of the centroids 
within a circular aperture of fixed radius. For 
Abell 665, 2163 and AC114 a 1 arcminute radius aperture was used. 
For Abell 2744, 773, 2218 and 2219 a 2 arcmin aperture was adopted,
and for Abell 520 a 3 arcmin aperture was used. For two of the clusters
included in the sample, Abell 665 and 1413, ROSAT HRI images were not
available at the time of writing and for these clusters PSPC imaging data have
been used instead. A summary of the ROSAT observations and the X-ray 
centers for the clusters is given in Table 3.

\subsection{Classification of clusters as cooling flow (CF) and 
non-cooling flow (NCF) systems}

For the purposes of this paper, we have classified the clusters into
subsamples of cooling-flow (hereafter CF) and non-cooling flow (NCF) 
systems. CFs
are those clusters for which the upper (90 percentile) limit to
the central cooling time, as determined from the deprojection analysis of
the ROSAT HRI X-ray images (Section 4), is less than $10^{10}$ yr. 
(The `central' cooling time is the mean cooling time of
the cluster gas in the innermost bin included in the deprojection
analysis, which is of variable size. The use of a fixed physical size of 
100 kpc for the central bin leads to similar results; Section 4.) 
Using this simple classification scheme we identify 21 CFs and 9
NCFs in our sample. The mean redshift for the subsamples of
both CF and NCF clusters is $\bar{z} = 0.21$.
The fraction of CFs in our sample is $70$ per cent,  
the same as that determined by Peres \etal (1998) 
from a study of a larger, flux-limited sample of clusters 
(primarily at lower redshift; $\bar{z} =0.056$) using a similar 
classification scheme.

\section{Spectral Analysis of the ASCA data}

\subsection{The spectral models }

The modeling of the X-ray spectra has been carried out using the XSPEC
spectral fitting package (version 9.0; Arnaud 1996). 
For the SIS data, only counts in pulse height analyser (PHA) channels
corresponding to energies between 0.6 and 10.0  \keV~  were included in the 
analysis (the energy range over which the calibration of the SIS 
instruments is best-understood). For the GIS data only counts in the energy 
range $1.0  - 10.0$ \keV~were used. The spectra were grouped before fitting to 
ensure a minimum of 20 counts per PHA channel, allowing $\chi^2$ statistics 
to be used.

The spectra have been modeled using the plasma codes of Kaastra \& Mewe
(1993; incorporating the Fe L calculations of Liedahl, Osterheld \&
Goldstein 1995, in XSPEC version 9.0) and the photoelectric absorption models 
of Balucinska-Church \& McCammon 
(1992). The data from all four ASCA 
detectors were analysed simultaneously (except where the S1 data were
lost due to chip saturation problems) with the fit parameters linked to 
take the same values across the data sets. The exceptions to this were the 
emission measures of the ambient cluster gas in the four detectors which, 
due to the different extraction radii used and residual uncertainties in the 
flux calibration of the instruments, were allowed to fit independently.

The spectra were examined with a series of spectral 
models. Model A, consisted of an isothermal plasma 
in collisional equilibrium, at the optically-determined
redshift for the cluster, and absorbed by
the nominal Galactic column density (Dickey \& Lockman 1990). 
The free parameters in this model were the temperature ($kT$) and 
metallicity ($Z$) of the plasma and the emission measures in the 
four detectors. (The metallicities are measured relative to
the solar values of Anders \& Grevesse (1989), with the different elements
assumed to be present in their solar ratios.) 
The second model, model B, was the same as model A but with the absorbing 
column 
density $(N_{\rm H})$ also included as a free parameter in the fits. 
The third model, model C, was the same as model A but with an 
additional component explicitly accounting for the emission from the 
cooling flows.  
The material in the cooling flows is assumed to cool at constant 
pressure from the ambient cluster temperature, following the prescription 
of Johnstone \etal (1992). The normalization of the cooling-flow 
component was parameterized in terms of a mass deposition rate, ${\dot
M_{\rm S}}$, which was a free parameter in the fits. The mass deposition
rate is linked to take the same value in all four detectors, scaled 
by a normalization factor proportional to the total 
flux measured in that detector. 
The metallicity of the cooling gas was assumed to
be equal to that of the ambient ICM. 
The emission from the cooling flows was also 
assumed to be absorbed by an intrinsic 
column density, $\Delta N_{\rm H}$, of cold gas 
which was a further free parameter in the fits. The abundances of
metals in the absorbing material were fixed to their solar values
(Anders \& Grevesse 1989) although the effects of varying these values were 
examined. Finally, a fourth spectral model, model D, 
was examined which was similar to model C but with the constant-pressure 
cooling flow replaced with a second isothermal emission component.
The temperature and normalization of this second emission component were
included as free 
parameters in the fits. (As with model C,  the normalizations were linked 
to the same value in all four detectors, scaled by the appropriate 
normalization factors.) The second emission component was also 
assumed to be intrinsically absorbed by a 
column density of cold gas, $\Delta N_{\rm H}$, which was a free parameter 
in the fits. The metallicities of the two emission 
components were linked to take the same values. 

Fig. 1 shows the ASCA data and best-fitting spectral models 
for four of the clusters in the sample; Abell 1704, 2029, 2204 and 2219.  
Table 4 summarizes the fit results for all of the clusters using the
four spectral models. For each cluster we list the best-fitting 
parameter values and 90 per cent ($\Delta \chi^2 = 2.71$) confidence
limits. The mass deposition rates (${\dot M_{\rm S}}$), $2-10$ keV X-ray 
luminosities ($L_{\rm X}$) and bolometric luminosities ($L_{\rm Bol}$)
are the values measured in the G3 detector. Note that for the clusters
at relatively 
low redshifts ($z<0.1$), we have corrected the $L_{\rm X}$ and $L_{\rm
Bol}$ values for the emission arising at radii $>6$ arcmin by scaling 
from the David \etal (1993) results, which were based on GINGA and Einstein 
MPC observations. 

\begin{figure*}
\hbox{
\hspace{0cm}\psfig{figure=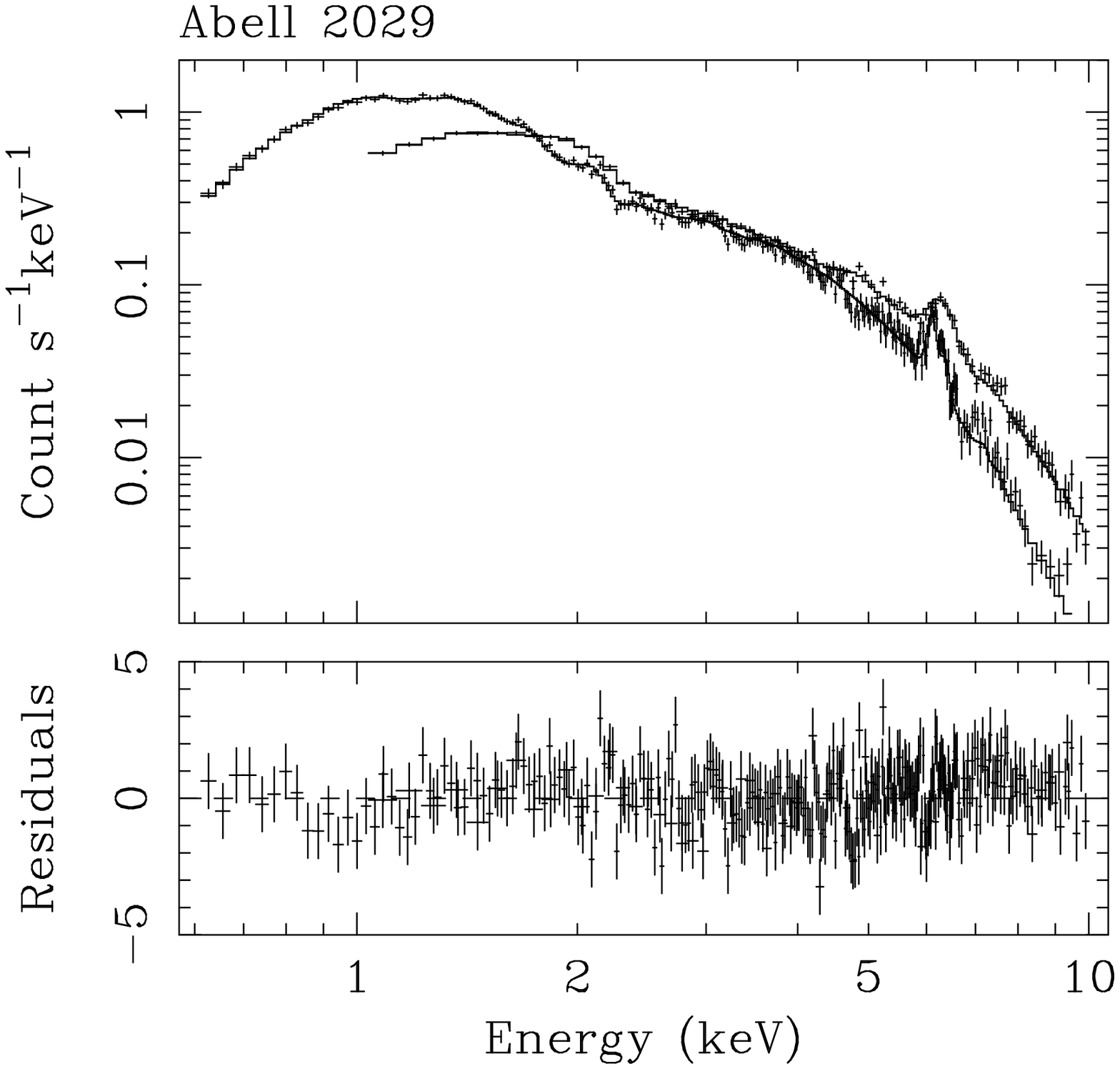,width=0.65\textwidth,angle=270}
\hspace{-2.5cm}\psfig{figure=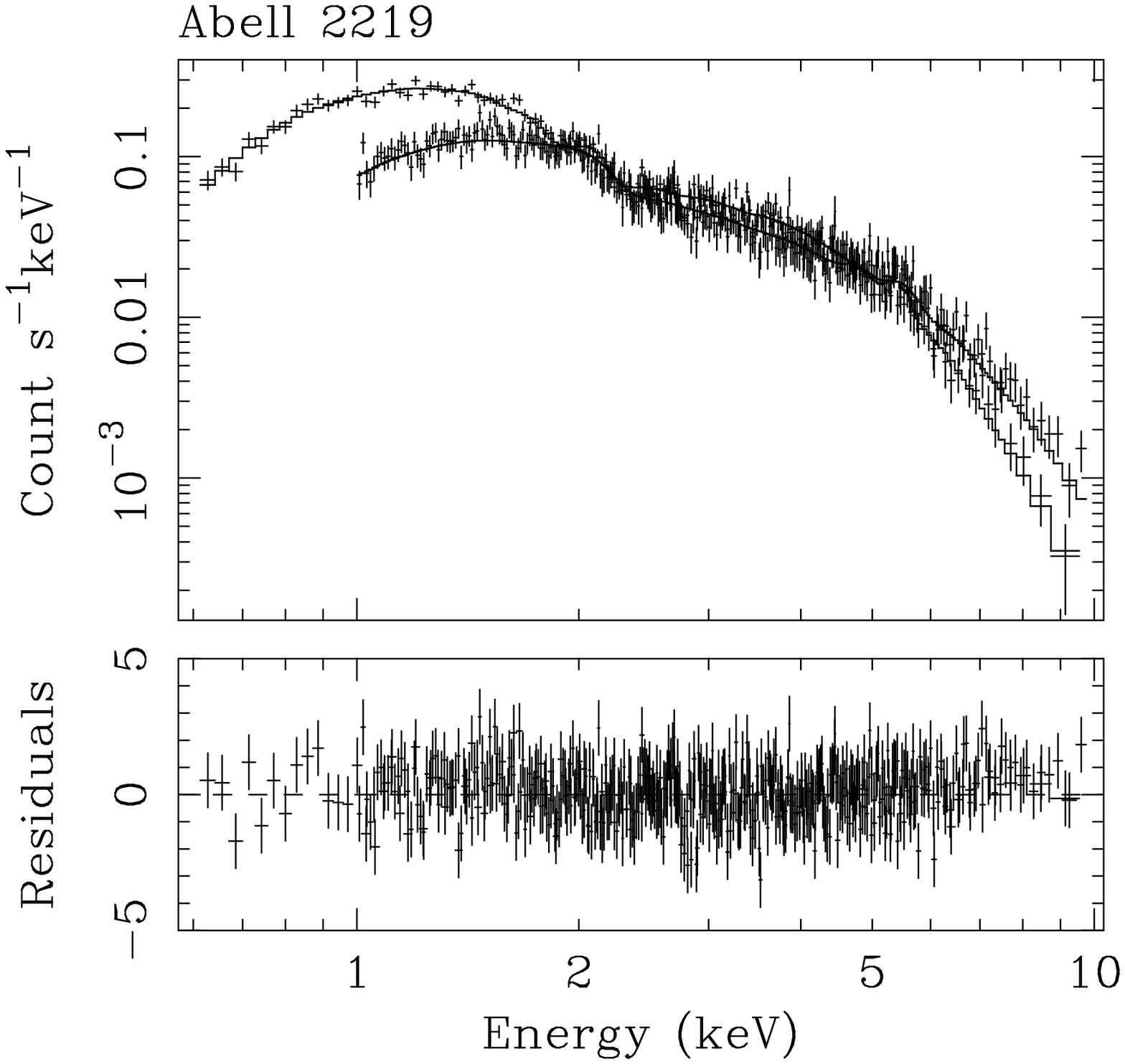,width=0.65\textwidth,angle=270}
}
\vspace{-0.2cm}
\hbox{
\hspace{0cm}\psfig{figure=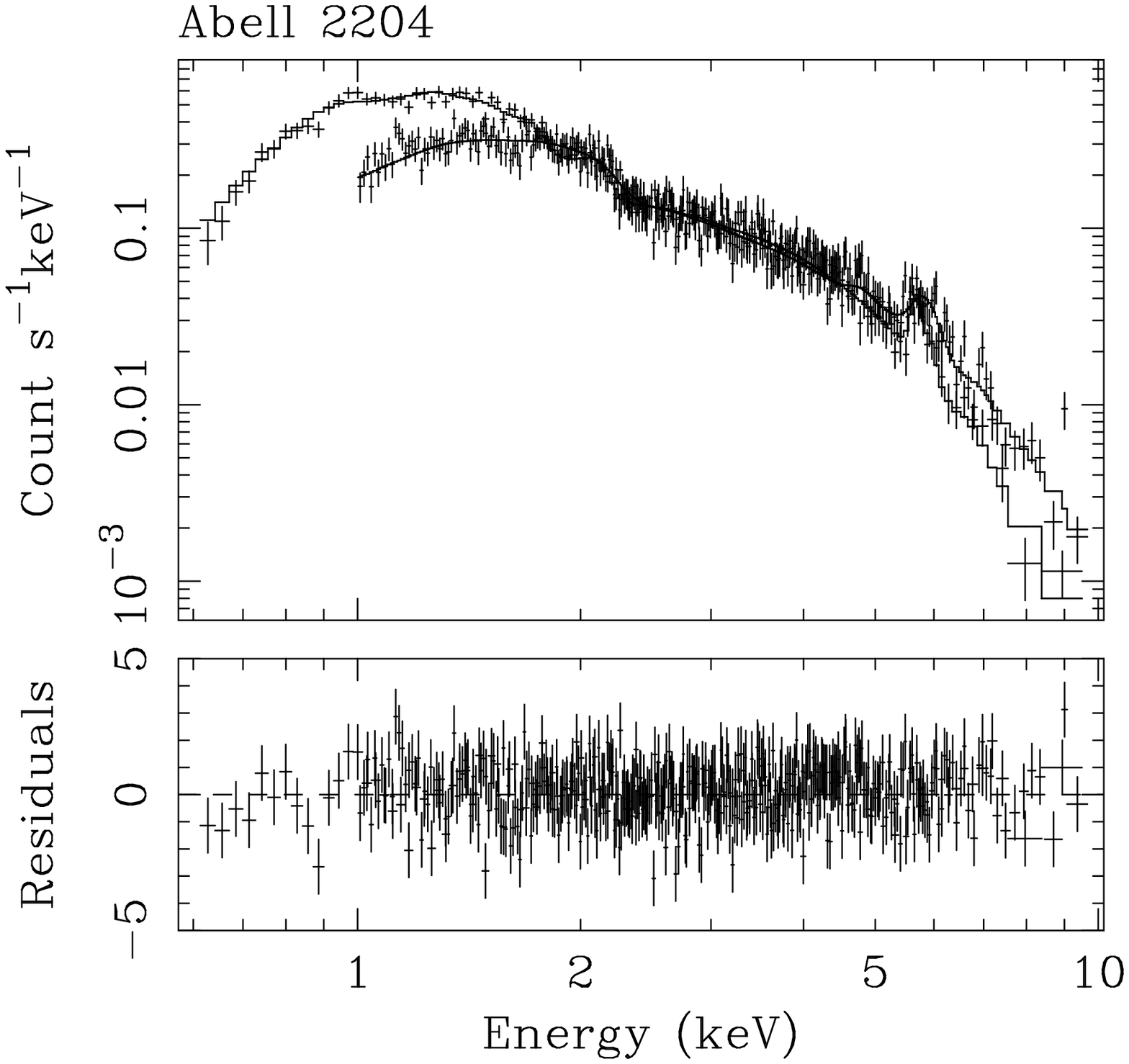,width=0.65\textwidth,angle=270}
\hspace{-2.5cm}\psfig{figure=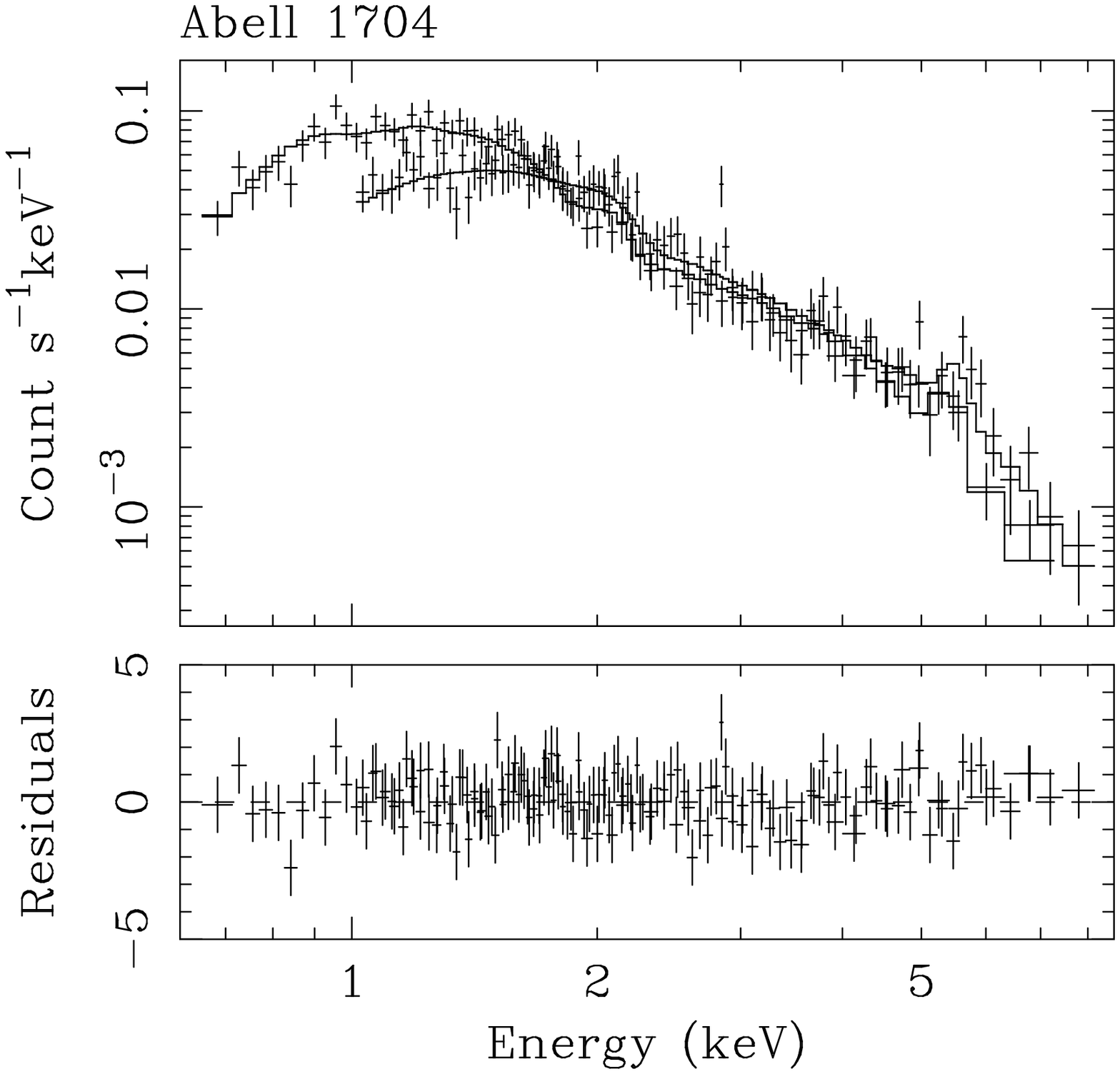,width=0.65\textwidth,angle=270}
}
\caption{(Upper panels) The ASCA data and best-fitting models 
and (lower panels) residuals to the fits (in units of $\chi$) 
for a representative subsample of the clusters included in the study. 
Clockwise, from top left, we show Abell 2029, 2219, 1704 and 2204.  
For the 3 CF clusters (Abell 2029, 2204 and 1704), the results 
determined with spectral model C are shown. For the NCF cluster, Abell 2219, 
spectral model B is shown. For clarity, only the S0 and G3 data are
plotted. The figure illustrates the range in quality of the ASCA data 
included in the study.} 
\end{figure*}

\subsection{The goodness-of-fit}

Table 5 summarizes the goodness-of-fit measurements 
obtained with the different spectral models. The tabulated
results are the probabilities of exceeding the $\chi^2$ values obtained, 
assuming in each case that the model correctly 
describes the spectral properties of the clusters. 
Goodness-of-fit values $< 10^{-2}$ may be regarded as indicating 
a formally unacceptable fit to the data. 

The results from the spectral analysis show that model A 
(the isothermal model with the absorbing column density fixed at the 
nominal Galactic value) provides a reasonable description of 
most (21/30) of the clusters. However, model A fails significantly 
for the massive CF clusters  Abell 1068, 1795, 2029 and 
2204. The significance of the improvements to the fits 
obtained with spectral model B over model A have been evaluated 
using the F-test for the introduction of an 
additional free parameter (Bevington 1969). 
These significances are listed in Column 6 of Table 5. 
We see that the improvements obtained when including the column density as 
a free parameter in the fits with the isothermal models 
are significant at $\geq 90$ per cent confidence for 23 of the 30 clusters 
(and would also be highly significant for Abell 478 if the nominal Dickey 
\& Lockman 1990 value for the Galactic column density, rather than the 
X-ray value determined by Allen \& Fabian 1997 were used in model A). 
Model B provides a statistically acceptable fit for 27 of the 30 clusters. 

The results on the goodness-of-fit determined with spectral models C and D 
are summarized in columns 4 and 5 of Table 5. The only clusters that these 
models do not adequately 
describe are the two nearest, brightest systems;  Abell 1795 and Abell 
2029. Spectral models C and D include an absorption component acting on  
the entire cluster emission which is normalized to the appropriate 
Galactic column densities determined from HI studies (Dickey \& Lockman 1990). 
We have examined the 
significance of the improvements to the fits obtained when 
allowing the Galactic column density to also be a free parameter in the 
fits with these models. 
The fits were not significantly improved for any of the
clusters except PKS0745-191,  for which the Galactic
column density is most uncertain. (The Galactic column density to
PKS0745-191 varies between 3.2 and $4.4 \times
10^{21}$ \apc~in the surrounding regions studied by Dickey \& Lockman
1990). We find $\Delta \chi^2 = 7.9$ with model C and 
$\Delta \chi^2 = 20.2$ with model D for PKS0745-191 (indicating improvements
significant at $>99$ per cent confidence) when allowing the Galactic column 
density to be free. The preferred value for the Galactic column density 
for this cluster from the ASCA data is 
$\sim 3.5 \times 10^{21}$ \apc. 
For Abell 1795, we also find a marginal improvement to the fits with
models C and D when allowing the Galactic column density to be a free
parameter ($\Delta \chi^2=4.5$ with 
Model C and $\Delta \chi^2=5.4$ for model D, which is significant at the 
96.5 and 98 per cent level, respectively) with a preferred value 
of $\leq 8 \times 10^{19}$ \apc. 
The agreement between the Galactic column densities inferred from the
21 cm and X-ray data for most of the clusters with spectral models C and D 
provides further support for the validity of these models.

\subsection{The requirement for multiphase models}

The statistical improvements obtained by introducing an additional 
emission component into the fits with the single-temperature  
model (model B) are summarized in Columns 7 and 8 of Table 5. We list the 
results obtained for both cases, where the extra emission component is 
modeled as a cooling flow (for the CF systems only) 
or a second isothermal emission component. (Essentially we use spectral 
models C and D, but also include the Galactic column density as 
a free parameter it the fits so as to permit a direct comparison with the 
goodness of fit measured with spectral model B). The significance of the 
improvements to the fits obtained with the multiphase models with respect to 
model B have been quantified using an F-test for the introduction of 2 or 3 
extra parameters (for models C and D respectively). We find that the 
improvements with model D are significant at $\geq 90$ per cent confidence 
for 21 of the 30 clusters. The only clusters for which we find no significant 
improvement with the introduction of a second emission component are the
NCF clusters Abell 520, 773, 2218 and AC114, and the more distant and/or 
less-luminous CF clusters Abell 963, MS1358.4+6245, MS1455.0+2232, 
MS2137.3-2353 and Abell 2390, which have amongst the lowest numbers of 
counts in their ASCA spectra. For the CF clusters (the only clusters 
to which model C was applied) the improvements obtained with the 
cooling-flow model over model B are significant at $\geq 90$ per cent 
confidence in 13 of 21 systems, with again those CF clusters with the lowest 
numbers of total counts in their ASCA spectra exhibiting the 
least-significant improvements.

We thus find that most CF clusters, and a number of NCF 
systems (in general those systems with the best signal-to-noise
ratios in their ASCA spectra) exhibit significant improvements to their
fits with the use of multiphase, over single-phase, models.  
These improvements generally indicate the presence of emission from 
material cooler than the mean cluster temperatures 
(see Table 4). For the CF clusters, this can be naturally understood as 
being due to emission associated with the cooling flows. 
For the NCF clusters, the improvements to the fits obtained
with model D are likely to reflect the presence of merging subclusters 
with lower virial temperatures. This is supported by optical, radio and X-ray
imaging (\eg Edge \etal 1992; Buote \& Tsai 1996; Markevitch 1996; 
Feretti, Giovannini \& B\"ohringer 1997; Rizza \etal 1998) and 
gravitational lensing studies (\eg Kneib \etal 1995; Smail \etal 1995, 
1997; Squires \etal 1997; Allen 1998) of the NCF clusters, which show
that these systems typically exhibit complex morphologies and centroid shifts 
indicative of merger events.

\subsection{Two-temperature versus cooling-flow models}

\begin{figure}
\centerline{\hspace{3cm}\psfig{figure=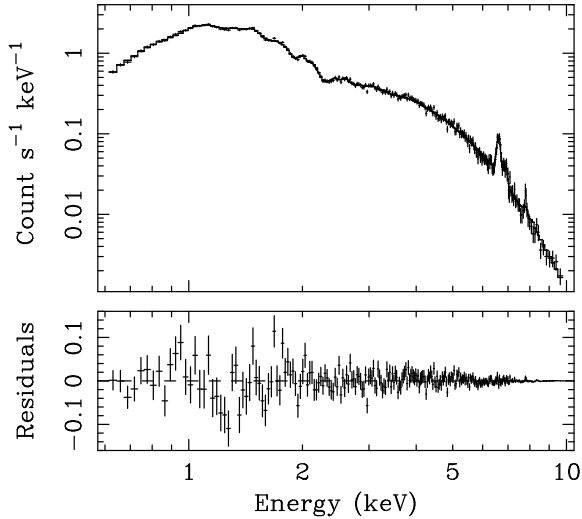,width=0.65\textwidth
,angle=270}}
\caption{(Upper panel) Simulated ASCA SIS spectrum of a 7 
keV cluster with a metallicity of 0.4$Z_{\odot}$ and a 
cooling flow, intrinsically absorbed by a column density of $4 \times 
10^{21}$\apc, accounting for 30 per cent of the total $2-10$ keV flux. 
A count rate of 1ct s$^{-1}$, an exposure time of 40ks and a Galactic 
column density of $10^{20}$\apc~have been assumed. The best-fitting 
two-temperature model (which provides a $\chi^2$ of 263 for 245 degrees of 
freedom) is overlaid. The best-fit parameters are $kT_1 = 6.7\pm0.4$ keV, 
$kT_2 = 1.4\pm0.2$ keV, $Z=0.41\pm0.04 Z_{\odot}$, with an 
intrinsic column density acting on the cooler component of $7.6\pm1.8 \times
10^{21}$ \apc). (Lower panel) The residuals to the fit in units of 
count s$^{-1}$keV$^{-1}$. }
\end{figure}

The results in Table 4 show that spectral model D typically provides at least 
as good a fit to the ASCA spectra for the CF clusters as the cooling-flow 
model (model C). This has sometimes been taken to indicate that the X-ray gas 
in these systems is distinctly two-phase, with the cooler phase being due to 
the dominant cluster galaxy (\eg Makishima 1997; Ikebe \etal 1999).
However, when interpreting these results it is important to recall that the 
two-temperature model provides a more flexible fitting parameterization, 
with an extra degree of freedom, when applied to ASCA observations. Simulated 
cooling-flow spectra constructed with spectral model C, including 
plausible levels of intrinsic absorption (Section 6) and observed at the 
spectral resolution and signal-to-noise levels typical of 
ASCA observations, are invariably well-described by two-temperature
models. This is illustrated in Fig. 2 where we show a simulated 
ASCA SIS spectrum for a 7 keV cluster with a metallicity of 0.4$Z_{\odot}$, 
containing a cooling flow (intrinsically absorbed by a column density of $4 
\times 10^{21}$\apc) accounting for 30 per cent of the total $2-10$ keV 
flux. A count rate of 1ct s$^{-1}$, an exposure time of 40ks and a Galactic 
column density of $10^{20}$\apc~have been assumed. Overlaid, we show the 
best-fitting two-temperature model, which provides a good fit to the 
simulated data. 

Thus, even in the case where the constant-pressure cooling flow model provides 
an exact description of the data ({\it i.e.} in the simulations), the 
two-temperature model provides a similarly 
good fit. In a real cluster, where spectral model C will undoubtedly 
over-simplify the true situation, the more flexible two-temperature model 
(model D) is likely to provide a better match to the observations. Thus, our
finding that spectral model D often provides a better fit to the observed
cluster spectra than model C, only shows that the constant pressure 
cooling flow model over-simplifies the true spectra of the cooling flows 
in the clusters (see also Section 6.4). We note that such tendencies 
would be enhanced by the presence of additional cool components in the 
spectra, for example due to temperature gradients at large radii 
(\eg Markevitch \etal 1998). Finally, we note that the very high 
luminosities associated with the cool emission components in clusters like 
RXJ1347.5-1145 (with $L_{\rm cool} \sim$ a few $10^{45}$\ergps) are difficult 
to explain with models in which the cool emission is due to the interstellar 
medium of the central cluster galaxies.

\section{Deprojection analysis of the X-ray images}

\subsection{Method and primary results}

The analysis of the HRI imaging data has been carried out using an 
extensively updated version of the deprojection code of Fabian \etal 
(1981). Azimuthally-averaged X-ray 
surface brightness profiles 
were determined for each cluster from the HRI images. 
These profiles were background-subtracted, 
corrected for telescope vignetting and re-binned 
to provide sufficient counts in each annulus to allow the 
analysis to be extended to radii of at least 500 kpc. 
(Bin sizes of 8-24 arcsec were used.)

With the X-ray surface brightness profiles as the primary input, 
and under assumptions of spherical symmetry and hydrostatic equilibrium, 
the deprojection technique can be used 
to study the basic properties of the intracluster gas 
(temperature, density, pressure, cooling rate) as a 
function of radius. The deprojection code uses a monte-carlo method to 
determine the statistical uncertainties on the results and 
incorporates the appropriate HRI spectral response matrix issued by GSFC. 
The cluster metallicities were fixed at the values 
determined from the spectral analysis (Table 4). The 
absorbing column densities were fixed at the appropriate 
Galactic values (Table 1).

The deprojection code requires the total 
mass profiles for the clusters (which define the
pressure profiles) to be specified. We have iteratively determined 
the mass profiles that result in deprojected
temperature profiles (which approximate the mass-weighted
temperature profiles in the clusters) that are isothermal within the regions 
probed by the HRI data (the central $0.5-1$ Mpc) and which are consistent 
with the spectrally-determined temperatures (Section 3). The assumption of 
approximately isothermal mass-weighted temperature profiles in the central regions 
of the clusters is supported by the following evidence: firstly, 
ASCA observations of nearby cooling flows show that 
in the central regions of these systems the gas is multiphase, but 
that the bulk of the X-ray gas there has a temperature close
to the cluster mean  (\eg Fabian \etal 1994b; Fukazawa \etal 1994;
Matsumoto \etal 1996; Ikebe \etal 1999). Secondly, combined X-ray and 
gravitational lensing studies of CF clusters (\eg Allen 1998) show that 
approximately isothermal mass-weighted temperature profiles in the cores of 
such clusters 
lead to excellent agreement between their X-ray and 
gravitational-lensing masses. Thirdly, the use of approximately constant 
mass-weighted temperature profiles implies a more plausible range of initial
density inhomogeneities in the clusters than would be the case if the 
temperature profiles decreased within their cores 
(Thomas, Fabian \& Nulsen 1987).  Finally, the use of approximately isothermal 
mass-weighted temperature profiles in the deprojection analyses 
leads to independent determinations of the mass 
deposition profiles in the cooling flows, 
from the X-ray spectra and imaging data, in excellent 
agreement with each other (\eg Allen \& Fabian 1997). 
We note that the assumption of a constant
mass-weighted deprojected temperature profile is
consistent with measurements of a decreasing emission-weighted 
temperatures in the cores of many CF clusters
(\eg Waxman \& Miralda-Escud\'e 1995). 

The mass profiles for the clusters were parameterized as isothermal spheres 
(Equation 4-125 of Binney \& Tremaine 
1987) with adjustable core radii ($r_{\rm c}$) and velocity dispersions
($\sigma$). The core radii were adjusted until the temperature profiles 
determined from the deprojection code became isothermal. The velocity 
dispersions were then adjusted until the temperatures determined from the 
deprojection code came into agreement with the spectrally-determined values. 
Errors on the velocity dispersions are the range of
values that result in isothermal deprojected temperature profiles 
that are consistent, at the 90 per cent confidence limit, with the 
spectrally-determined temperature results. Estimates of the thermal gas 
pressure in the outermost radial bins used in the analysis are also required 
by the deprojection code and were determined iteratively.  
(The uncertainties on the outer pressure estimates do not significantly affect 
the results presented here.) Although the deprojection method of Fabian 
\etal (1981) is essentially a single-phase technique, we note that it 
produces results in good agreement with more detailed multiphase treatments 
(Thomas, Fabian \& Nulsen 1987) and, due to its simple applicability at 
large radii in clusters, is better-suited to the  present project.

The mass distributions determined from the deprojection analysis 
are summarized in columns 3 and 4 of Table 6.  These simple
parameterizations permit direct comparisons with independent mass
constraints from gravitational lensing and dynamical studies (\eg Allen
1998). We note that for a few of the clusters (primarily the 
brightest CFs with the best data) the single-component mass models cannot 
adequately satisfy the requirement for approximately isothermal
deprojected temperatures profiles and for these systems
a significantly better match was obtained by introducing a second `linear' 
mass component, truncated at a specified outer radius. The clusters requiring the additional 
mass components were Abell 586 ($2 \times 10^{11}$ \Msun kpc$^{-1}$ within 
the central 20 kpc), Abell 2029 ($3 \times 10^{11}$ \Msun kpc$^{-1}$
within the central 20 kpc), Abell 2219 ($10^{11}$ \Msun kpc$^{-1}$ within the 
central 30 kpc), Abell 478 ($4 \times 10^{11}$ \Msun kpc$^{-1}$ within
the central 20 kpc and an adjustment of the cluster parameters to 
$\sigma = 840^{+80}_{-60}$\kmps~and  $r_{\rm c} = 100$ kpc) and 
Abell 1795 ($10^{11}$ \Msun kpc$^{-1}$ within the 
central 20 kpc, and an adjustment of the cluster parameters to 
$\sigma = 720^{+20}_{-20}$\kmps~and  $r_{\rm c} = 60$ kpc).

The core radii of the mass distributions determined from the deprojection 
analysis are similar to the values measured from 
simple `$\beta-$model' fits to the X-ray surface brightness profiles  
(\eg Cavaliere \& Fusco-Femiano 1976, 1978). It is well known that the 
$\beta$-model does not provide a good match to the X-ray surface 
brightness profiles of the brightest CF clusters when their 
central regions are included in the analysis (\eg Jones \& Forman 1984). 
However, where $\beta$-models are used to estimate the mass core radii 
in CF clusters, the central regions of these clusters should not 
simply be excised from the fits. The central surface brightness peaks in 
CF clusters trace the presence of sharp central density rises in 
these systems and are not only the result of excess emission due to 
cooling gas. (It is the high central densities that lead to the high 
cooling rates.) Although fully accounting for the effects of 
cooling on the central density measurements is difficult with current 
data, simply excising the inner regions of the clusters can lead to
overestimates of the mass core radii and corresponding underestimates of 
the central gas and total mass densities. 

The basic results on the cooling flows from the deprojection analysis
are summarized in columns $5-8$ of Table 6. For those clusters in common,
the results obtained are generally in good agreement with values reported 
from previous works (\eg Edge \etal 1992, White \etal 1997; Peres 
\etal 1998).

\section{The mass deposition rates from the cooling flows}

\begin{figure}
\centerline{\hspace{3cm}\psfig{figure=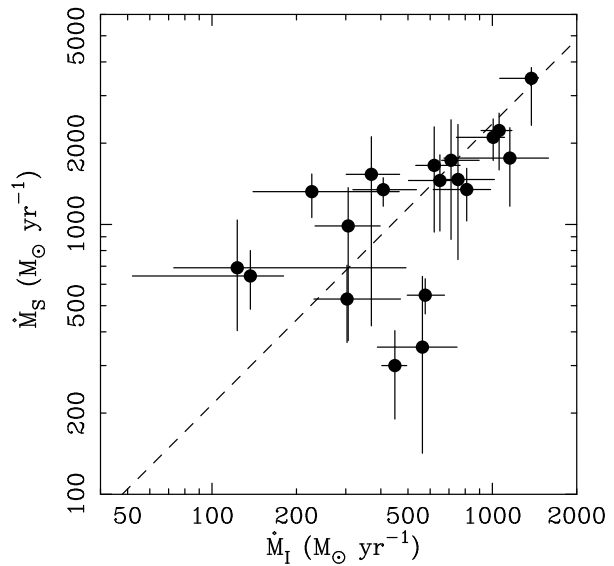,width=0.65\textwidth
,angle=270}}
\caption{The mass deposition rates determined from the 
ASCA spectra (${\dot M_{\rm S}}$) versus the values measured from the 
deprojection analysis of the
ROSAT HRI imaging data (${\dot M_{\rm I}}$). The dashed line is the
best-fitting power-law model. No correction for the effects of intrinsic
absorption on the deprojection results has been made. }
\end{figure}

\begin{figure}
\centerline{\hspace{3cm}\psfig{figure=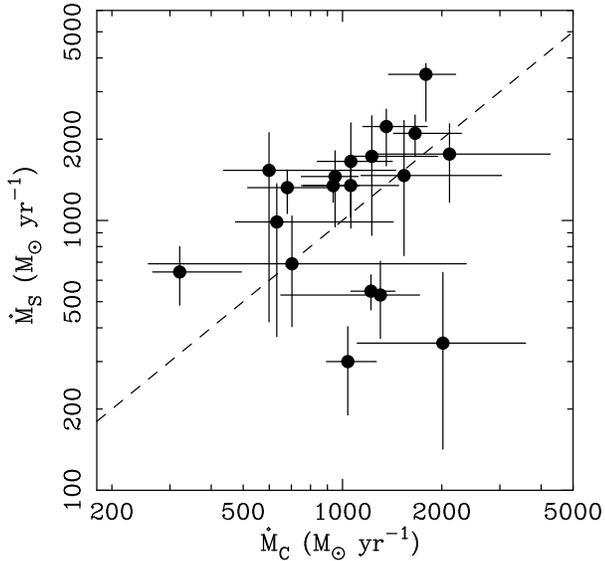,width=0.65\textwidth
,angle=270}}
\caption{The mass deposition rates determined from the 
ASCA spectra (${\dot M_{\rm S}}$) versus the values measured from the 
deprojection analysis, corrected for 
the effects of intrinsic absorption due to cold gas (${\dot M_{\rm C}}$). 
The dashed line indicates equality between the values.
}
\end{figure}

\subsection{Comparison of the spectral and imaging results}

The spectral and image deprojection analyses discussed 
in Sections 3 and 4
provide essentially independent estimates of the mass deposition rates in 
the clusters. A comparison of the results obtained from these analyses
therefore provides a test of the validity of the cooling flow
model. 

 The deprojection method describes the X-ray emission from 
a cluster as arising from a series of concentric spherical shells. 
The luminosity in a particular shell, $j$, may be written as  the sum of 
four components (Arnaud 1988).

\begin{equation}
L_j = \Delta{\dot M}_jH_j + \Delta{\dot M}_j \Delta \Phi_j +
\left[\sum_{i=1}^{j-1} {\Delta{\dot M}_i (\Delta \Phi_j + \Delta H_j
})\right],
\label{swa_eq1}
\end{equation}

\noindent where $\Delta{\dot M}_j$ is the mass deposited in shell $j$,
$H_j$ is the enthalpy of the gas in shell $j$, and $\Delta \Phi_j$ is
the gravitational energy released in crossing that shell. $\sum_{i=1}^{j-1}
{\Delta{\dot M}_i }$ is the mass flow rate through shell $j$, and $\Delta
H_j$ the change of enthalpy of the gas as it moves through that shell.
The first term in equation \ref{swa_eq1} thus accounts for the enthalpy of
the gas deposited in shell $j$. The second term is the gravitational work
done on the gas deposited in shell $j$. The third and fourth terms
respectively account for the gravitational work done on material flowing 
through shell $j$ to interior radii, and the enthalpy released by that 
material as it passes  through the shell. 

In any particular shell, the densest material in the cooling flow is assumed 
to  cool out and be deposited. Since the cooling time of this 
material will be short compared to the flow time, the cooling
can be assumed to take place at a fixed radius. Thus, the luminosity 
contributed by the first term in equation 1 should have a spectrum 
that can be approximated by gas cooling at constant pressure from the ambient 
cluster temperature \ie the same spectrum as the cooling component 
incorporated into the spectral analysis with model C (Section 3).

For the bulk of the material continuing to flow inwards towards the
cluster centre, the cooling via X-ray emission is assumed to be offset by 
the gravitational work done on the gas as it moves inwards. The emission 
accounted for in the second and third terms of equation 1 should therefore 
have a spectrum that can be approximated by an isothermal plasma at the appropriate ambient 
temperature for the cluster \ie the spectrum of the isothermal emission 
component in model C.  Since the mass-weighted temperature profiles in 
the clusters are assumed to remain approximately isothermal with radius, 
the luminosity contributed by the fourth term of equation \ref{swa_eq1} 
should be negligible. 

The mass deposition rates, ${\dot M_{\rm I}}$, listed in Table 6 are the 
mass flow rates ($\sum_{i=1}^{j-1} {\Delta{\dot M}_i }$) determined from the
deprojection analysis, at the point 
where the mean cooling time of the cluster gas first exceeds the Hubble
time ($1.3 \times 10^{10}$ yr).  Thus, if the cooling flow model is correct, 
and the cooling flows have been undisturbed for a significant fraction of 
a Hubble time, the mass deposition rates determined from the deprojection 
analysis should be similar to those measured independently from the 
spectral data.

Fig. 3 shows the mass deposition rates determined from the spectral
analysis (${\dot M_{\rm S}}$) versus the image deprojection results 
(${\dot M_{\rm I}}$), for those clusters for which
both quantities have been measured. We see that the results exhibit
an approximately linear correlation; a fit to the data
with a power-law model of the form ${\dot M_{\rm S}} = P {\dot 
M_{\rm I}^Q}$, using the Akritas \& Bershady (1996) bisector
modification of the ordinary least-squares statistic, gives a 
best-fitting power-law slope of $1.0\pm 0.2$ (where the error is the
standard deviation determined by bootstrap re-sampling), although
the spectrally-determined mass deposition rates typically exceed the 
deprojection results by a factor of $2-3$. However, the spectral
analysis presented in Section 3 also requires that the cooling gas is 
intrinsically absorbed by equivalent hydrogen column densities of, typically,
a few $10^{21}$ \apc~(model C). 
The mass deposition rates 
determined from the deprojection analysis must therefore also 
be corrected for the effects of this absorbing material. 

The corrections for the effects of intrinsic absorption on the 
deprojection results have been carried out by re-running the
deprojection analysis with the absorbing column densities set to the total 
values determined with spectral model C. We assume that the absorption is 
due to cold gas with solar metallicity. If, however, the intrinsic absorption 
were instead due to dust, a possibility examined in more detail in 
Section 7, then the required correction factors would likely be reduced 
by a few tens of per cent. (The introduction of a simple OIK absorption 
edge at $E\sim 0.54$keV, such as might be associated with oxygen-rich, 
silicate dust grains, generally
provides as good a description of the intrinsic absorption 
detected in the $0.6-10.0$keV ASCA spectra as a cold, gaseous absorber.)
Since the intrinsic column densities measured with model C 
are redshifted quantities, we set the total column densities
used in the revised deprojection analysis, which assumes zero redshift 
for the absorber, to be $N_{\rm H} + \Delta N_{\rm H}/(1+z)^3$, where
$N_{\rm H}$ is the Galactic column density). 
The absorption-corrected mass deposition rates (${\dot M_{\rm C}}$) 
so determined are summarized in Table 7. Fig. 4 compares the 
spectrally-determined mass deposition rates with the absorption-corrected 
deprojected values. 
We see that once the presence of intrinsic absorption has been 
accounted for in a consistent manner in the imaging and spectral
analyses, the results on the mass deposition rates in the clusters 
show good agreement.

\subsection{The fraction of flux from the cooling flows}

Table 8 lists the fractions of the total $2-10$ keV X-ray fluxes from 
the CF clusters associated with cooling gas, as determined with spectral 
model C. The results range from $\sim 10$ per cent for systems like Abell 
1689 and 2142 to $>30$ per cent for Zwicky 3146, Abell 1068, RXJ1347-1145 
and Abell 2204. Such a range of measurements is consistent with the results 
of Peres \etal (1998), from a deprojection analysis of a flux-limited sample 
of the brightest clusters known.

The fraction of the X-ray flux from a cluster due to it's cooling
flow can be expected to increase with the age of the flow. It is therefore 
likely that the cooling flows in Abell 1689 and 2142 are younger 
than those in Abell 2204 and Zwicky 3146. We note that the X-ray data 
are not, in general, of sufficient quality to place firm constraints on  
the ages of the cooling flows in these clusters (for a discussion 
of cooling flow age measurements see Allen \etal 1999). 

It is interesting that the clusters with the largest cooling-flow 
flux fractions (with the exception of Abell 1413 and 2261) are also amongst the 
most optically line-luminous cooling flows known (\eg Crawford 
\etal 1999). The production of powerful optical emission lines and 
associated UV/blue continuum emission is known to be due, at least 
in part, to the formation of massive stars at the centres of cooling flows 
(\eg Johnstone, Fabian \& Nulsen 1987; Allen 1995; Cardiel \etal 1995, 1998; 
McNamara \etal 1996; Voit \& Donahue 1997). The results presented here 
are consistent with the idea that such star formation is typical of 
undisturbed 
cooling flows. It is also interesting to note that the 
clusters with the largest cooling flow flux fractions and most optically 
line-luminous emission line nebulae also tend to have the shortest cooling times 
averaged over their central 100 kpc regions, with $\overline{t_{\rm 100}} 
\approxlt 3$Gyr (see also Peres \etal 1998).

\section{Intrinsic X-ray absorption in clusters}

\subsection{The measured column densities and model dependency of the 
results} 

\begin{figure*}
\hbox{
\hspace{0cm}\psfig{figure=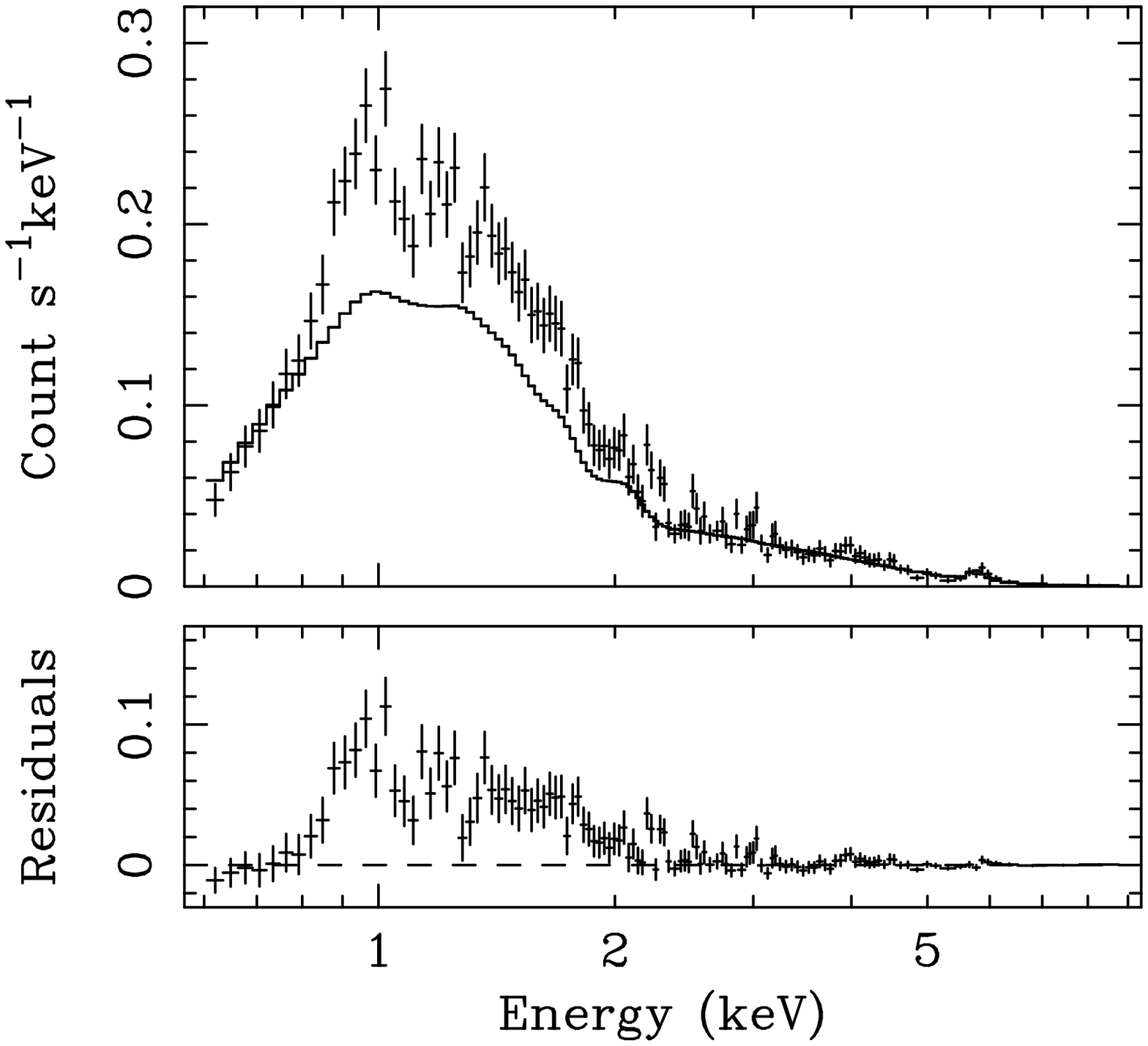,width=0.65\textwidth,angle=270}
\hspace{-2.5cm}\psfig{figure=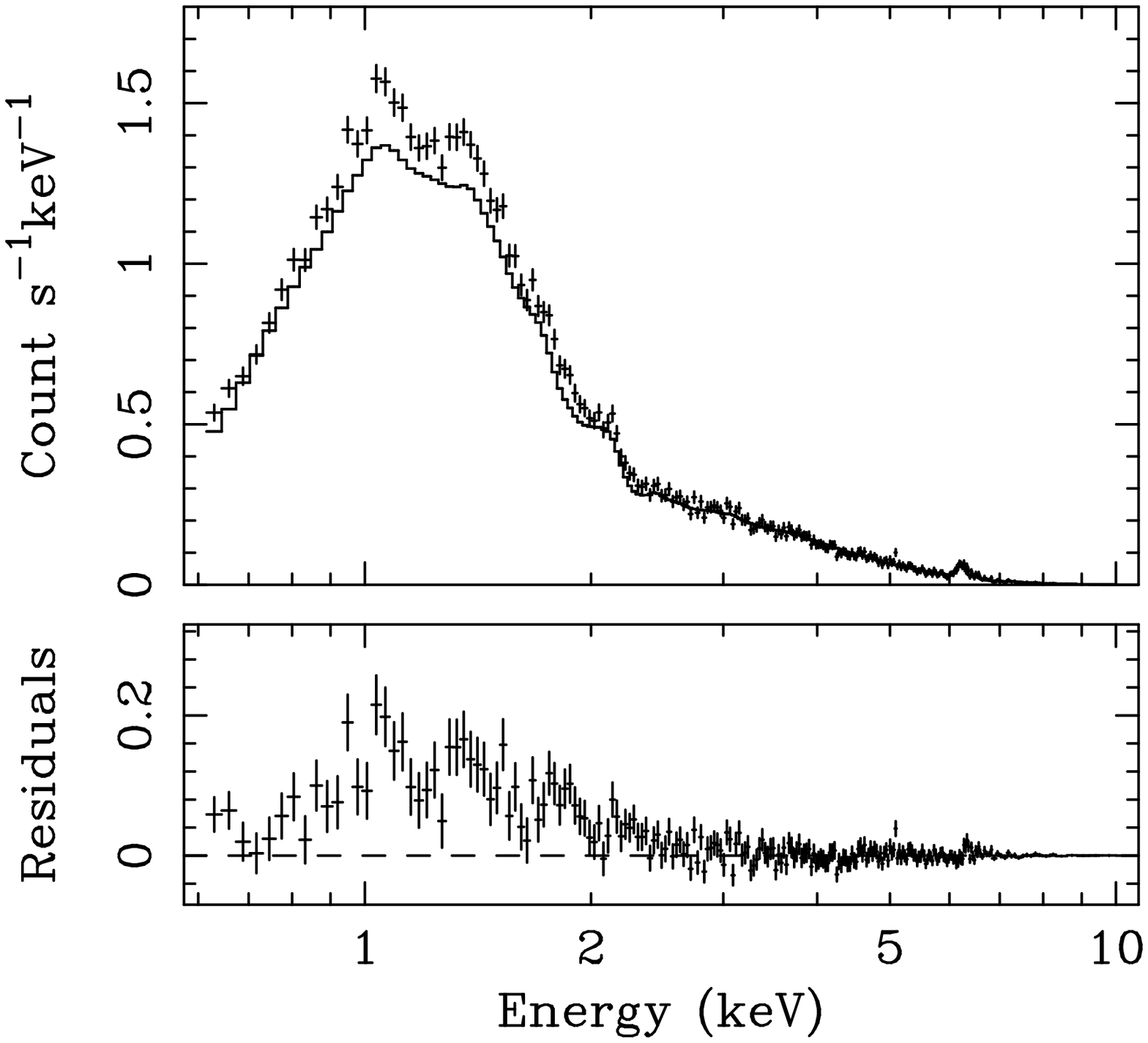,width=0.65\textwidth,angle=270}
}
\caption{The data, best-fit models and residuals from the fits to the 
SIS0 spectra for Abell 1068 and 1795 in the energy range $3.0-10.0$ keV, 
using spectral model A. The best-fit models have then been extrapolated to 
cover the full $0.6-10.0$ keV band. The upper panels show the data and 
best-fit models. The lower panels show the residuals to the fits. The 
figures show that the residuals, which can be naturally explained by the 
introduction of cool emission components with associated intrinsic
absorption, take the form of an excess of counts between 
energies of $\sim 0.8-2.5$ keV.}
\end{figure*}

The equivalent column densities of intrinsic X-ray absorbing material inferred 
from the spectral analysis are summarized in
Table 9. All three of the spectral models incorporating a free-fitting 
absorption component (models B,C,D) reveal the presence of excess X-ray 
absorption over and above the nominal Galactic values along the lines of
sight to the clusters (Dickey \& Lockman 1990), although the measured
column densities are sensitive to the spectral model used. 

Using the simple isothermal model with free-fitting absorption (model B),
we determine a mean excess column density for the whole sample of 30
clusters of $3.8 \pm 3.2 \times 10^{20}$ \apc.  
When we examine the CF and NCF clusters separately, we find
that both subsamples exhibit intrinsic absorption, at a similar
level on average. For the CF clusters, the mean excess column density 
measured with spectral model B is  $3.5 \pm 3.0 \times 10^{20}$ \apc. For 
the NCF systems, the value is $4.4 \pm 3.9 \times 10^{20}$ \apc. 
The application of a Students t-test (accounting for the possibility of 
unequal variances in the two distributions; Press \etal 1992) shows the 
mean excess column densities for the CF and NCF clusters, determined with
spectral model B, to differ at only the 45 per cent confidence level.
The application of a Kolmogorov-Smirnov test shows the two subsamples to be 
drawn from different parent populations at only the $\sim 10$ per cent confidence
level. Thus, our analysis with spectral model B suggests that intrinsic
absorption is not confined to CF clusters. 

The results on the column densities are quite different, however, when the 
more-sophisticated multiphase spectral models are used. As discussed in 
Section 3, the multiphase models generally provide a better description of the 
X-ray properties of the CF clusters (and some of the brighter NCF systems). 
Using spectral model C, our preferred model for the CF clusters in that it
provides a consistent description of the spectral and imaging X-ray data
for these systems (Section 5.1), we determine a mean intrinsic column 
density acting on the cooling-flow components of $3.4 \pm 1.3 \times 
10^{21}$ \apc. Using the two-temperature spectral model (model D), we 
determine a mean intrinsic column density acting on the cooler emission 
components in the CF systems of $6.3 \pm 4.7 \times 10^{21}$ \apc. Thus, the 
intrinsic column densities for the CF clusters measured with the multiphase 
models (models C and D) are similar and approximately an order of magnitude 
larger than the values inferred using the single-phase model B. These results 
demonstrate the need for adopting an appropriate spectral when attempting
to measure the column densities of absorbing material in CF clusters. 
We also note that where spectral model D provided a significant improvement
to the fits to the NCF clusters, with respect to model B,  the measured
column densities acting on the cooler emission components were also
significantly larger than the values determined using the single-phase models. 

We have also measured the intrinsic column densities in the 
CF clusters using one further spectral model, referred to in Table 9 as 
model C'. Model C' is identical to model C except that it assumes that
the excess absorption acts on the entire cluster spectrum, rather than
just the cooling gas, and that the absorbing material lies at zero redshift. 
Model C' has been used in a number of previous studies and is included here 
for comparison purposes. 
The mean excess column density for the CF clusters determined with spectral 
model C' is $7.2 \pm 3.9 \times 10^{20}$ \apc.

Allen \& Fabian (1997) present results from an X-ray colour 
deprojection study of 18 clusters observed with the ROSAT PSPC. 
These authors determine intrinsic column densities across 
the central 30 arcsec (radius) regions of Abell 478, 1795 and 2029,  
using spectral model C', of $7.40\pm0.63$, 
$2.20 \pm 0.20$ and $0.83 \pm 0.42 \times 10^{20}$ \apc, respectively.
The results for Abell 478 and Abell 1795 are in excellent agreement with those 
presented here, although the Allen \& Fabian (1997) value for Abell 2029
is $\sim 4$ times smaller than our result. Abell 2029 is unusual in that it 
hosts a strong cooling flow without associated optical line emission. 
This may indicate that the central regions of the cluster have been 
disrupted (perhaps by a minor merger event or the strong central radio source 
in this system), complicating the distributions of cooling and absorbing 
material ({\it cf.} Section 5.2).  

Four of the CF clusters studied here were also examined by
White \etal (1991), using Einstein Observatory SSS data. 
These authors measured intrinsic column densities 
for Abell 478, 1795, 2029 and 2142 (in a 3 arcmin radius circular aperture, using spectral model C') 
of $1.7^{+0.8}_{-0.7} \times 10^{21}$ \apc, $0.8^{+0.3}_{-0.3} \times 
10^{21}$ \apc, $1.8^{+0.5}_{-0.5} 
\times 10^{21}$ \apc and $1.3^{+0.3}_{-0.4} \times 10^{21}$ \apc, 
respectively. (The SSS result for Abell 478 has been corrected to 
account for the different value of Galactic absorption assumed in that
study). The intrinsic column densities determined  
by White \etal (1991) are $2-5$ times larger than the values 
measured from the ASCA data. 

Allen \& Fabian (1997) compared the results from their X-ray colour
deprojection study of PSPC data with the White \etal (1991) SSS 
analysis and concluded that if the intrinsic absorption were due to cold gas, 
then agreement between the measured column densities could only be 
obtained if the covering fraction of the X-ray absorbing material
were, in general, $\approxlt 0.5$. We have therefore examined the constraints 
on the covering fraction, $f$, that can be obtained from the ASCA data 
using spectral model C. The results are also listed in Table 9. 
In all cases we determine a best-fit covering fraction for the intrinsic 
X-ray absorbing material of unity. For many clusters, covering
fractions of $< 70$ per cent can be firmly ruled out. (The physical
significance of these results are further explored in Section 6.4). 
The results on the covering fractions, and the reasonable agreement of 
the ASCA and 
PSPC results for Abell 478 and 1795, suggest that the White \etal 
(1991) results may have systematically over-estimated the intrinsic column 
densities in clusters by a factor of a few. 

Before considering in more detail the possible origin and nature of 
the absorbing material, it is pertinent to consider whether the intrinsic
absorption, inferred to be present using a variety of spectral
models, could in fact be an artifact due to an unconsidered emission 
process in the clusters. Fig. 5 shows the SIS0 spectra for Abell 1068 
and 1795, which exhibit two of the clearest absorption signatures. 
The data between $3.0$ and $10.0$ keV have been fitted with a 
single-temperature emission model with Galactic absorption ({\it i.e.}
spectral model A). The residuals to these fits, over the full $0.6-10.0$
keV band, are shown in the lower panels. We see that the residuals, 
which can be naturally explained by the introduction of cool
emission components with associated intrinsic absorption 
(using spectral models C and D), take the form of a large excess of 
counts between $\sim 0.8-2.5$ keV. We therefore include 
the caveat that if some extra emission process which can account for such
residuals in the X-ray spectra were identified, then the requirement for 
intrinsic absorption from the X-ray data would be greatly diminished. 
Finally, we note that the fits with spectral Model A over the restricted 
$3.0-10.0$ energy range lead to determinations of the cluster temperatures 
for Abell 1068 and 1795 of $6.1^{+2.6}_{-1.6}$ and $6.1^{+0.5}_{-0.4}$ keV,
respectively, in excellent agreement with the results obtained 
for the full data sets with spectral model C.

\subsection{The mass of absorbing material}

The results presented in Section 6.1 show that the cooling flows in our
sample are typically intrinsically absorbed by equivalent hydrogen 
column densities of a few 
$\times 10^{21}$ \apc~(taking spectral model C as our preferred model).
Assuming, in the first case, 
that this absorption is due to cold gas with solar metallicity, the
mass of absorbing gas, $M_{\rm abs}$, within a radius, $r_{\rm abs}$, 
may be approximated as 

\begin{equation}
M_{\rm abs} \sim 3.2 \times 10^7 r_{\rm abs}^2 \Delta N_{\rm H} \Msun,
\label{swa_eq3}
\end{equation}

\noindent where $r_{\rm abs}$ is in units of kpc and $\Delta N_{\rm H}$ in
$10^{21}$ \apc. For metallicities of $0.4 Z_\odot$~in the absorbing gas 
(the mean 
emission-weighted value for the X-ray emitting gas in the CF clusters; 
Allen \& Fabian 1998b) the implied masses are a factor $\sim 2$ larger. 
For zero metallicity these masses are a factor $\sim 3$ times larger than 
the values given by equation 2. (The
scaling factors for different metallicities in the absorbing material have 
determined from spectral simulations using the XSPEC code.) 

We have adopted $r_{\rm abs}$ as the radius where the cooling time 
of the cluster gas first exceeds $5 \times 10^9$ yr in the clusters (a
plausible  age for the cooling flows in most of the CF systems). 
These radii, and the masses of absorbing gas implied by equation 2 (using
the column densities determined with spectral model C) for both solar 
metallicity and $Z=0.4Z_\odot$, are listed in columns $2-4$ of Table 10.
 
The observed masses of absorbing material may be compared to the masses
expected to have been accumulated by the cooling flows $(M_{\rm acc})$ 
within the same radii over the past $5 \times 10^9$ yr. Assuming that 
from time $t=0$ to $t=5 \times 10^9$ yr, the integrated mass deposition rate 
within radius $r_{\rm abs}$ increases approximately linearly with time, the
total mass accumulated within radius $r_{\rm abs}$ after $5 \times 10^9$ 
yr is approximately 

\begin{equation}
M_{\rm acc} \sim 2.5 {\dot M}(r<r_{\rm abs}) \times 10^{9} \Msun.
\label{swa_eq4}
\end{equation}

At times $t > 5 \times 10^9$ yr, the mass deposition rate
within $r_{\rm abs}$ should remain approximately constant. If instead
we assume ${\dot M}(r<r_{\rm abs})$ to be constant during the first 5 Gyr,
the accumulated mass will be twice the value indicated by equation
3. (Exactly how the mass deposition rate within $r_{\rm abs}$ grows 
with time during the first 5 Gyr is unclear and will depend 
upon the evolution of the cluster and the detailed properties of the cluster
gas.) 

The mass deposition rates within radii $r_{\rm abs}$ (corrected for the
effects of absorption due to cold gas, as discussed in Section 5.1) and 
the estimated 
masses of cooled material accumulated by the cooling flows within these radii 
over a 5 Gyr period, are summarized in Table 10. Fig. 6 compares the
accumulated masses with the masses of absorbing material
determined from the spectral data. The agreement between the results is
reasonable and is maximized for a metallicity in the absorbing gas of 
$\sim 0.4-0.6Z_{\odot}$ (depending on the growth of the cooling flow 
over the first 5Gyr). This supports the idea that the observed X-ray 
absorption may be due to material 
accumulated by the cooling flows. We note that if the X-ray absorption 
in the CF clusters were due to dust rather than cold gas, a possibility 
examined in more detail in Section 7, then the agreement between the 
observed and predicted masses of absorbing material shown in Fig. 6 
might still be expected to hold if the dust were contained in the 
material deposited by the cooling flows. In this case, the  
mass of the absorber calculated with equation 2 would be the mass of 
cooled gas associated with the dust when it was deposited from the 
cooling flow. For a Galactic dust to gas ratio, the mass in dust  
would be $\sim 100$ times smaller than the associated gas mass.

\begin{figure*}
\hbox{
\hspace{0cm}\psfig{figure=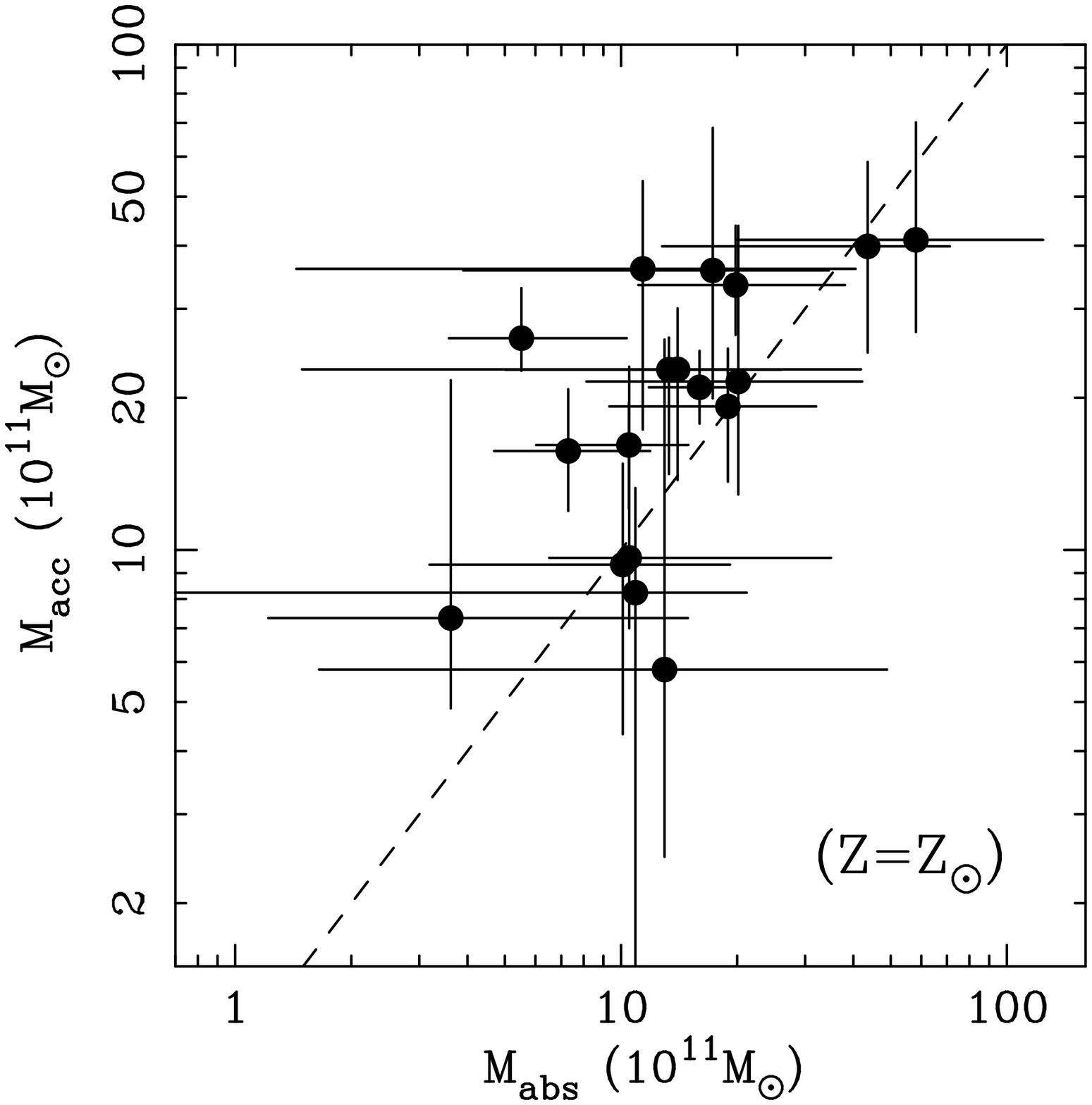,width=0.65\textwidth,angle=270}
\hspace{-2.5cm}\psfig{figure=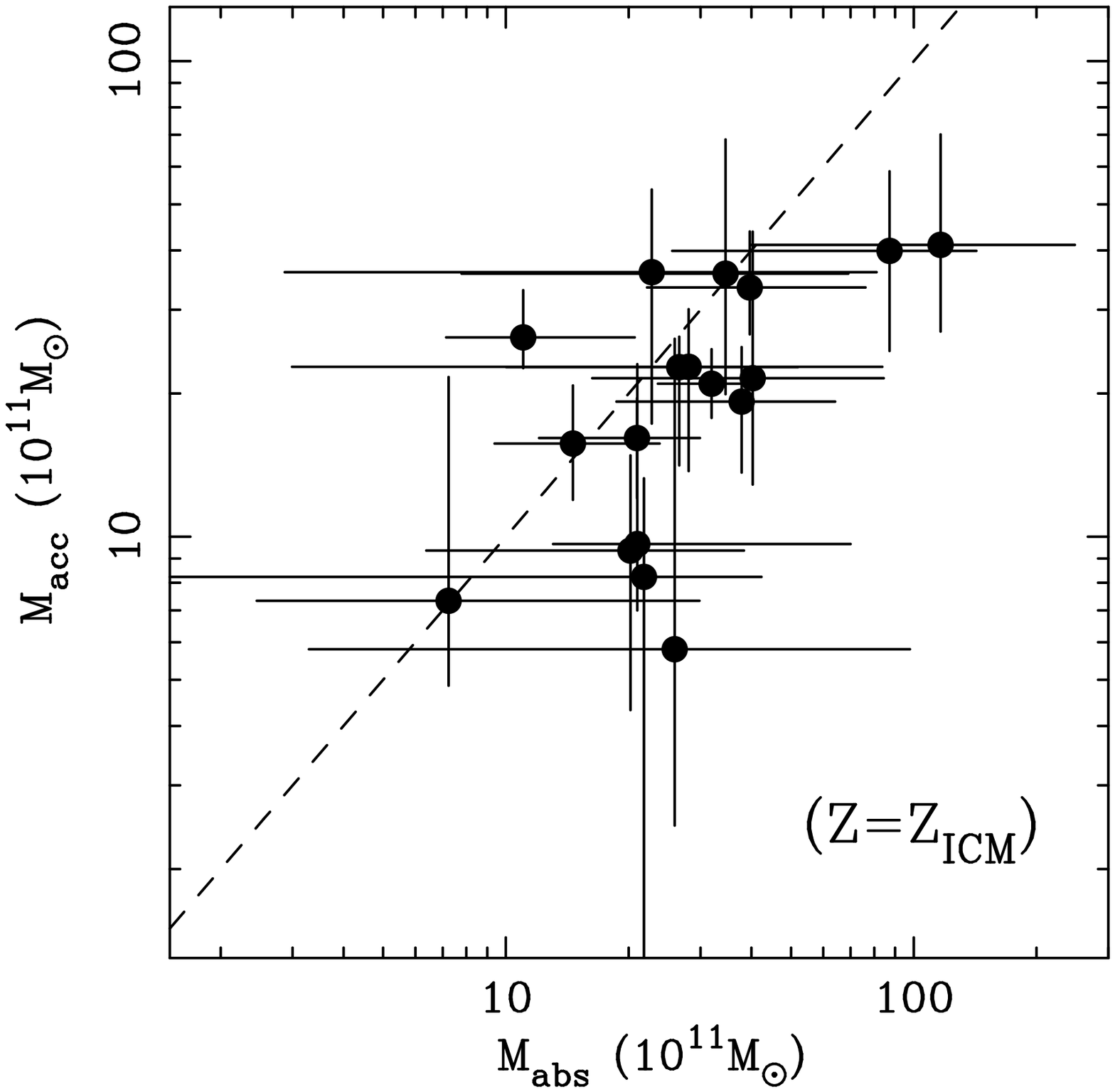,width=0.65\textwidth,angle=270}
}
\caption{A comparison of the masses predicted to have been 
accumulated by the cooling flows within radii 
$r_{\rm abs}$ over the first 5 Gyr, versus the masses within the same radii 
implied by the observed intrinsic X-ray absorption (assuming
the absorption to be due to cold gas with a covering fraction of unity). 
The dashed line indicates equality between the two sets of values. 
Results are shown both for (left) solar metallicity in the absorbing gas 
and (right) a metallicity of $0.4 Z_{\odot}$.}
\end{figure*}

\subsection{The luminosity reprocessed in other wavebands}

The luminosities absorbed at X-ray wavelengths must eventually be
reprocessed in other wavebands. If the absorbing material is dusty 
(as is likely to be the case in the central regions of the clusters 
\eg Voit \& Donahue 1995; Fabian, Johnstone \& Daines 1994; Allen \etal 1995) 
then the bulk of this reprocessed emission is likely to emerge in the 
infrared band (\eg Dwek, Rephaeli \& Mather 1990). Table 11 summarizes 
the reprocessed luminosities ({\it i.e.} the bolometric luminosities 
absorbed within the clusters) measured with spectral model C. The 
reprocessed luminosities range from $\sim 2 \times 10^{44}$ \ergps~for 
Abell 1795 to $\sim 5 \times 10^{45}$ \ergps for RXJ1347.5-1145, with a mean 
value of $\sim 10^{45}$ \ergps. 

Allen \etal (1999) report detections of spatially 
extended $100\mu$m emission coincident with the X-ray centroids 
in the nearby Centaurus cluster and Abell 2199. The infrared fluxes from those 
clusters were also shown to be in good agreement with the predicted values 
due to reprocessed X-ray emission from their cooling flows. We have also 
used the IPAC SCANPI software and archival 
IRAS data to measure the 60 and $100\mu$m fluxes within a four arcmin (radius) 
aperture centred on the X-ray centroids for the clusters in the present
sample. (The medians of the co-added SCANPI results were used.) The results are listed 
in  Table 11. The error bars associated with 
the measured fluxes are the root-mean-square deviations in the residuals, external 
to the source extraction regions, after baseline subtraction. Where no
detection was made, an upper limit equal to three times the r.m.s. 
deviation in the residuals is listed.  Where a detection was
made, we also list the in-scan separations (in arcmin) between the 
peak of the $100\mu$m emission (or the $60\mu$m emission in the case of
Abell 1835) and the X-ray centre.

Two of the clusters in our sample, IRAS 09104+4109 and Abell 1068 
have known infrared point sources coincident with their X-ray centroids 
(Kleinmann \etal 1988; Moshir \etal 1989). The presence of these
sources is confirmed by our analysis. The 60 and $100\mu$m data for 
Abell 1704 and $60\mu$m data for Abell 1835 also suggest the presence 
of unresolved sources, coincident with the X-ray centroids for the clusters. 
Abell 478 and Zwicky 3146 have $100\mu$m emission originating from close to 
their X-ray centroids, which appears spatially extended, although 
the infrared flux from Abell 478 is probably contaminated by 
Galactic cirrus. The data for Abell 2142, 2204, 2261 and 2390 
also provide significant detections within the four arcmin (radius) source 
apertures, although the peaks of the infrared emission from these systems 
are spatially offset from their X-ray centroids, suggesting that the detected 
flux is likely to originate, at least in part, from some other source. 

Several of the clusters included in our study (Abell 586, 963, 1413,
1795, 2029, 2142 and PKS0745-191) have previously been studied at 
infrared wavelengths, using IRAS data, by Wise \etal (1993) and/or Cox, 
Bregman \& Schombert (1995). These authors also report no clear 
detections of infrared emission from these sources. Edge \etal (1999) 
report measurements of $60\mu$m emission from Abell 1835 and 
60 and $100\mu$m emission from Abell 2390, using a similar analysis
to that presented here. These authors also present detections 
of $850\mu$m emission from these clusters which, for Abell 1835, 
they suggest is likely to be due to dust heated by the vigorous 
star formation observed in the cluster core or an 
obscured active galactic nucleus.  

Following Helou \etal (1988) and Wise \etal (1993), we can estimate 
the total infrared luminosities implied by the observed IRAS fluxes and flux
limits using the relation 

\begin{equation}
L_{\rm 1-1000\mu m} \sim 2.8 \times 10^{44} (\frac{z}{0.05})^2(2.58S_{60}+S_{100}) \ergps,
\end{equation}

\noindent where $S_{60}$ and $S_{100}$ are the 60 and $100\mu$m IRAS
fluxes in units of Jy. This relation assumes a dust temperature of $\sim
30$K (which is consistent with the observations; \eg Allen \etal 1999; 
Edge \etal 1999) and an emissivity 
index, $n$, in the range $0-2$, where the emissivity is proportional to the 
frequency, $\nu^n$. Where measurements, rather than upper limits, were
obtained, we associate a systematic uncertainty of $30$ per cent with the 
estimated $1-1000\mu$m luminosities, which is combined in quadrature with 
the random errors. The $1-1000\mu$m luminosities calculated from this relation 
are summarized in Table 11. In general, the measurements and upper limits 
to the observed $1-1000\mu$m luminosities exceed the predicted 
luminosities due to reprocessed X-ray radiation from the cooling flows. 
Where measurements (rather than upper limits) are made, this may suggest
the presence of an additional source of infrared flux, such as star 
formation or AGN associated with the central cluster galaxies, or 
some other contaminating (possibly Galactic) source. At some level, 
the massive starbursts at the centres of systems like Abell 1068, 1835, 2204 
2390 and Zwicky 3146 (Allen 1995; Crawford \etal 1999; Edge \etal 1999) 
must contribute to the detected infrared flux.

\subsection{Systematic uncertainties in the absorption measurements}

As discussed in Section 6.1, the measured intrinsic equivalent hydrogen 
column densities of absorbing material inferred from
the ASCA spectra exhibit significant systematic variation, depending upon 
whether the single-phase or
multiphase emission models are used. We have adopted the results 
obtained with the cooling-flow model (model C) as our preferred values, 
since this model 
provides a consistent description of the spectral and imaging X-ray 
data for the clusters. However, a number of systematic uncertainties 
affecting the absorption results remain with the cooling flow model.

Spectral model C undoubtedly over-simplifies the situation in a real
cooling flow. In particular, the model assumes that the absorbing material 
lies in a  uniform screen in front of the cooling flows. If the absorber is 
intrinsically associated with the clusters, then the absorbing 
material  must, to some extent, be 
distributed throughout the cooling flows. Allen \& Fabian (1997; see also
Wise \& Sarazin 1999) discuss two different, limiting geometries for the 
absorbing material; partial covering and multilayer absorption. The partial covering model, with a
covering factor $f<1$, can be expected to apply or where there are $<1$ 
individual 
absorbing clouds along each line of sight and/or where the absorbing material 
is strongly clumped 
on large scales. However, as discussed in Section 6.1, covering fractions 
significantly less than unity are found to provide a poor description of the 
ASCA spectra,
suggesting that the absorbing material is likely to be 
more smoothly distributed throughout the cooling flows.  

The multilayer absorption model discussed by Allen \& Fabian (1997)
is applicable where the absorbing material is made up of a large
number of similarly sized absorbing clouds, each with a column density much 
less than the total intrinsic column density ($\Delta N_{\rm H}$), 
homogeneously distributed throughout the X-ray emitting medium. 
For an intrinsic emission spectrum $A(E)$, the observed spectrum, $A'(E)$, 
emerging from the emitting/absorbing region may be written (where $\sigma(E)$ 
is the absorption cross-section) as

\begin{equation}
A'(E) = A(E) \left ( \frac {1 - e^{-\sigma(E) \Delta N_{\rm H}}}
{\sigma(E) \Delta N_{\rm H}} \right ).
\end{equation}

We have simulated cooling flow spectra with multilayer absorption, with
total column densities of between $10^{21}$ and  $5\times10^{22}$ \apc, and 
fitted these with simple models which assume that the absorbing 
material acts as a uniform screen in front of the cooling flow 
(as with model C) . For the simulations we adopt an emission spectrum appropriate for a 
constant pressure cooling flow, with an upper temperature of 7 keV and a 
metallicity of $0.4Z_{\odot}$. We use a response matrix appropriate for 
the ASCA SIS detectors. The comparison between the fitted (uniform screen) 
column densities and the true total (multilayer) values are shown in Fig. 7. 
We see that for multilayer column densities of $10^{21}-5\times 10^{22}$\apc, 
the fitted column densities underestimate the true values by factors of 
$5-10$. (These results are not sensitive to any reasonable choice of Galactic 
column density.) 

If the multilayer model approximates the true
distribution of X-ray emitting and absorbing material in cooling flows,
then the results obtained with spectral model C would imply true equivalent 
column-densities of intrinsic absorbing gas in the cluster cores of,
typically, a few $10^{22}$\apc. Importantly, we also note that for 
signal-to-noise ratios appropriate for ASCA observations of bright clusters 
(with count rates of $\sim 1$ct s$^{-1}$ in the SIS detectors and exposure 
times of $\sim 40$ ks), the simple uniform-screen absorption model 
provides a good fit to the simulated spectra with multilayer absorption, for 
total column densities $\approxlt 3\times10^{22}$ \apc. (For larger total 
column densities, complex residuals are detected in the spectra at 
energies $\approxlt 2.0$keV.) The large intrinsic column densities implied by 
the multilayer models 
would have strong implications for the nature of the absorbing matter. 
In particular, column densities of $\approxgt 10^{22}$\apc~of gaseous 
material with solar metallicity would be 
very difficult to reconcile with current 
HI and CO limits, even if this material were very cold and highly molecular. 
(We also note that in the case of a multilayer distribution, the implied
masses of absorbing gas would often exceed the predicted accumulated 
masses due to the cooling flows and imply, at least in part, some other
origin for the absorbing medium.)
In the following section, we explore these issues in more detail
and summarize the current observational constraints on the physical nature
of the absorbing matter.

\begin{figure}
\centerline{\hspace{3cm}\psfig{figure=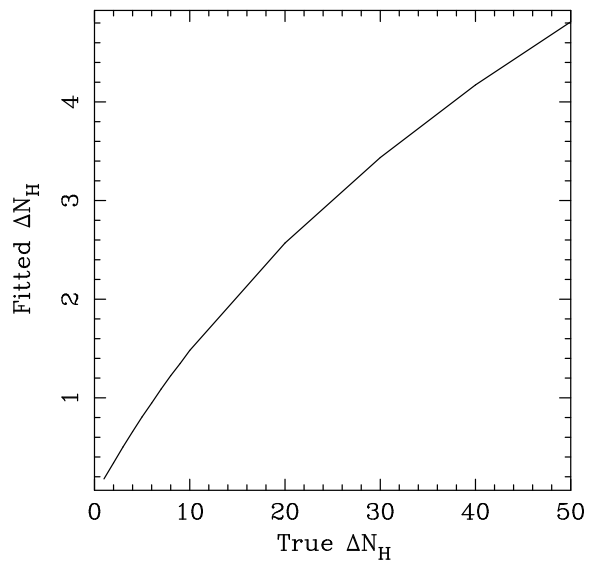,width=0.65\textwidth
,angle=270}}
\caption{The apparent column density determined from fits to cooling flow 
spectra with a uniform-screen absorption model (\eg spectral model C)
when in reality the absorber is distributed in a multilayer manner 
(Allen \& Fabian 1997). The vertical axis shows the fitted column density 
and the horizontal axis 
the true total (multilayer) column density through the emitting/absorbing 
region. Both axes are in units of $10^{21}$ \apc.} 
\end{figure}

\section{Searches for the X-ray absorbing material in other wavebands}

The results presented in this paper provide a 
consistent picture for the X-ray properties of cooling flows in 
X-ray luminous clusters of galaxies. The ASCA spectra and ROSAT 
images provide independent determinations of the mass 
cooling rates in the clusters in good agreement with each 
other, when a consistent method of analysis is employed (Section 5.1). 
The spectral data require the presence of large column densities of 
intrinsic X-ray absorbing material associated with the cooling gas 
(Section 6). The implied masses of absorbing material are in reasonable 
agreement with the masses expected to have been accumulated by the cooling 
flows over their lifetimes (Section 6.2), identifying this material as a 
plausible sink for the bulk of the cooled gas from the flows. Despite the 
evidence for intrinsic X-ray absorption associated with cooling flows, 
however, searches for this material in other wavebands have, to date, 
proved largely unsuccessful (at least outside of the central $5-20$ kpc). We 
here briefly review these observations and comment on
their implications for the physical state of the X-ray absorbing matter. 
More detailed discussions of a  number of these issues are given by Ferland, 
Fabian \& Johnstone (1994), Daines, Fabian \& Thomas (1994), Fabian \etal
(1994a),  O'Dea \etal (1994a), 
Voit \& Donahue (1995), O'Dea \& Baum (1996), Arnaud \& Mushotzky (1988)
 and Henkel \& Wiklind (1998).

As mentioned in Section 5.2, the central ($5-20$ kpc) regions 
of CF clusters are commonly observed to be sites of ongoing star formation 
with powerful, associated optical line emission 
(\eg Johnstone \etal 1987; Heckman \etal 1989; 
McNamara \& O'Connell 1989; Crawford \& Fabian 1992; 
Allen 1995; Cardiel \etal 1995, 1998; McNamara \etal 1996; Voit \& Donahue
1997; Crawford \etal 1999). Nearly all CF clusters with short central 
cooling times ($t_{\rm cool} \approxlt 2$Gyr) exhibit such phenomena 
(Peres \etal 1998), whereas systems with longer central cooling times 
(including all NCF clusters) generally do not. The observed levels of 
star formation imply the presence of at least moderate masses of cooled gas in the 
cores of CF clusters, sufficient in a few cases to account for the 
bulk of the mass deposited by the cooling flows in those regions 
(Allen 1995) and at least some causal connection with the cooling flows. 

At radio wavelengths, extensive searches have been carried out 
for 21cm line-emission associated with warm atomic hydrogen 
in the cores of CF clusters (\eg Valentijn \& Giovanelli 1982;
O'Dea \& Baum 1996). The negative results obtained (with the exception of
M87; Jaffe 1992) typically constrain the column densities of optically thin 
material to be less than a few $10^{19}$ \apc~and, for an 
assumed spin temperature, $T_{\rm S} \sim 20$K, the number of clouds 
along the line of sight to be $\approxlt 1$ (O'Dea \& Baum 1996).

Searches have also been made for the 21 cm absorption signature of 
cold atomic hydrogen (\eg McNamara, Bregman \& O'Connell 1990;
Jaffe 1992; Dwarakanath, van Gorkom \& Owen 1994; O'Dea, Gallimore 
\& Baum 1995; O'Dea \& Baum 1996; Johnstone \etal 1998). 
Such material has been detected across the central few kpc of a 
few CF clusters, including Abell 426 (Crane, Van der Hulst \& Haschick
1982; Sijbring \etal 1989; Jaffe 1990) 2A0335+096 
(McNamara \etal 1990), Hydra A (Dwarakanath, van Gorkom \& Owen 1995), MKW3s
(McNamara \etal 1990) and Abell 2597 (O'Dea, Baum \& Gallimore 1994) with 
typical column densities of a few $10^{20}(T_{\rm S}/100)$
\apc. However, for most clusters only negative results have been obtained, 
with upper limits to the column densities of 10K gas 
(assuming a covering fraction of unity and a velocity width of 
$\sim 1.5$ kms$^{-1}$;  Dwarakanath \etal 1994, O'Dea \& Baum 1996) of 
$10^{18}-10^{19}$ \apc.  

Laor (1997) presents tight limits on the column density of atomic 
hydrogen towards the core of Abell 426 ($\Delta N_{\rm H} \approxlt 4 
\times 10^{17}$ \apc) from Ly$\alpha$ observations made with the Hubble Space 
Telescope (see also Johnstone \& Fabian 1995). Together, the results 
on HI emission and absorption, and Ly$\alpha$ absorption, imply that the bulk 
of the material responsible for the observed X-ray absorption cannot be in the 
form of atomic hydrogen.

An issue of crucial importance in interpreting the 21cm HI results and the 
results from CO studies (see below) is 
the expected temperature of X-ray absorber, if in gaseous form, which has 
proved controversial. Johnstone, Fabian \& Taylor (1998; updating earlier 
calculations by Ferland, Fabian \& Johnstone 1994 and Fabian \etal 1994a) suggest that cooled gas clouds, in equilibrium with the 
cooling-flow environment, are likely to be very cold. 
Johnstone \etal (1998) suggest that dust-free clouds
form an extended outer envelope, with a column density of $\sim 5 
\times 10^{21}$ \apc and a temperature of 13-17K, beyond which the gas drops 
to the microwave background 
temperature. For Galactic dust/gas ratios, however, the results are
significantly modified, such that the cloud temperature drops to the 
microwave background temperature after a column density of only 
$\sim 10^{20}$ \apc. At such low temperatures, dusty gas is likely to be 
highly molecular. However, these predictions differ from 
those of O'Dea \etal (1994a) and Voit \& Donahue (1995) who suggest a 
minimum temperature for such clouds of $\sim 20$K, even in the presence 
of dust. Braine \etal (1995) also suggest a minimum temperature of $\sim
10$K for the absorbing gas. Possible reasons for the discrepancies
between these results (with contrasting views) are discussed by Voit \& 
Donahue (1995) and Ferland \etal (in preparation).

Extensive searches have been carried out for CO 
associated with molecular gas in cooling flows. To date CO emission has 
only been detected in Abell 426 (Lazareff \etal 1989; Mirabel, Sanders
D.B. \&  Kaz\'es I. 1989). Braine \etal (1995) also present limits on CO absorption in the central regions 
of this cluster). The upper limits to the column density of molecular hydrogen 
in the inner regions of other clusters (for an assumed 
kinetic temperature of $T_{\rm CO}^{\rm K}=20$K and a covering fraction 
of unity)  are typically $\approxlt$ a  few $10^{20}$ \apc (O'Dea \etal
1994a, 
Antonucci \& Barvainis 1994, McNamara \& Jaffe 1994; Braine \& Dupraz 1995; 
O'Dea \& Baum 1996) which is significantly below the X-ray inferred column 
densities. The CO results also constrain the number of clouds along the line 
of sight (for $T_{\rm CO}^{\rm K}=20$K) to be $\approxlt 10$ 
(O'Dea \etal 1994a; O'Dea \& Baum 1996). The CO emission limits appear 
to exclude the possibility of large column densities ({\it i.e.} 
values consistent with the X-ray measurements) of molecular gas, 
with a kinetic temperature $\approxgt 10$K. However, if the bulk of the 
X-ray gas cools to the microwave background temperature, as suggested by 
Johnstone \etal (1998), the CO limits may be consistent with the 
X-ray data. 

It should be noted that the determinations of molecular hydrogen masses from 
CO observations are affected by a number of additional uncertainties. 
Firstly, the CO/H$_2$ ratios in cluster cooling flows may differ from 
Galactic values. In particular, where the temperature and metallicity of the 
absorbing gas is low, the use of a standard Galactic CO/H$_2$ ratio may 
underestimate the mass of molecular gas (Maloney \& 
Black 1988; Madden \etal 1997). 

It is unclear whether X-ray absorbing gas accumulated by cooling flows 
could have 
substantially sub-solar metallicity. Allen \& Fabian 
(1998b) determine a mean emission-weighted metallicity for the X-ray 
emitting gas in CF clusters of $Z \sim 0.4Z_\odot$. These authors 
also show that CF clusters generally contain metallicity gradients, so
that the metallicity of the X-ray gas within the cooling radii of CF 
systems will, on average, exceed this value. Thus, it seems unlikely that 
the material currently being deposited from the cooling flows will have 
substantially sub-solar metallicities. However, if the X-ray absorbing 
material is formed from the matter deposited by the cooling flows over
their lifetimes, the metallicity of this material could be somewhat 
lower. Within an inhomogeneous cooling flow (\eg Nulsen 1986, 1998; Thomas, 
Fabian \& Nulsen 1987) the denser gas at any particular radius will
tend to have a smaller volume filling factor. If the metals are evenly 
distributed throughout the volume of the cooling flow at any particular 
radius (whether this will occur is unclear), the least 
dense gas will contain most of the metals (Reisenegger, Miralda-Escud\'e \& Waxman 1996). 
Since the densest material in cooling flows will have the shortest cooling 
time and be deposited first, the material accumulated by the cooling flows
over their histories could have a significantly lower metallicity 
than the material being deposited today. Conversely, if the 
metals in cooling flows were concentrated in the densest material, 
the metallicity of the X-ray absorbing material could exceed that 
of the X-ray emitting gas. 

Fabian \etal (1994a), have suggested that the masses 
of molecular gas inferred from CO observations could be 
significantly underestimated if most of the CO 
in the absorbing material has frozen onto dust grains. Such freezing 
should occur rapidly (on timescales $\sim 10^5$ yr) if even small amounts 
of dust are initially present in the absorbing gas (Voit \& Donahue 1995).
Daines \etal (1994) and Fabian \etal (1994a) discuss
how dust is likely to form even in clouds that are initially free 
of dust, on timescales $\sim 10^9$ yr (although see also Voit \& Donahue
1995). Optical, UV and sub-mm studies of the cores of cooling flows 
show that dust lanes (\eg Sparks, Macchetto \& Golombek 1989; 
McNamara \& O'Connell 1992) and intrinsic reddening 
(Hu 1992; Allen \etal 1995; Edge \etal 1999) are common. 

Elston \& Maloney (1994) and Jaffe \& Bremer (1997)
report detections of H$_2$(1-0)S(1) emission from warm ($\sim 2000$K) 
molecular hydrogen from K-band spectroscopy of the inner few kpc of a number 
of CF systems (including Abell 478 and and PKS0745-191). 
No detections of such emission have been made in NCF systems. 
The H$_2$S emission is likely to arise from the heated skin of cold 
molecular clouds which could also be responsible for the optical 
emission-line phenomena discussed above. 

\begin{figure*}
\hbox{
\hspace{0cm}\psfig{figure=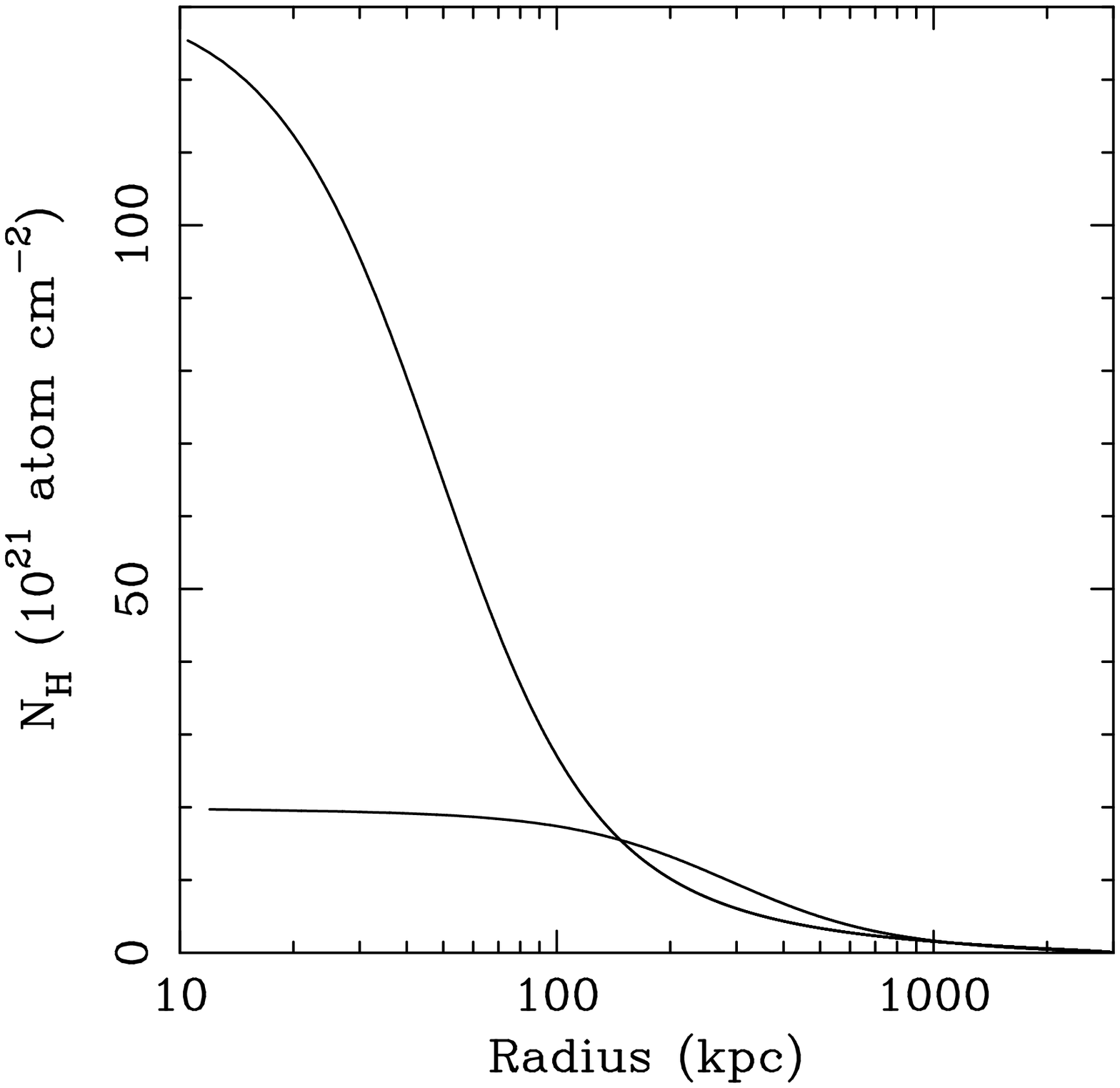,width=0.65\textwidth,angle=270}
\hspace{-2.5cm}\psfig{figure=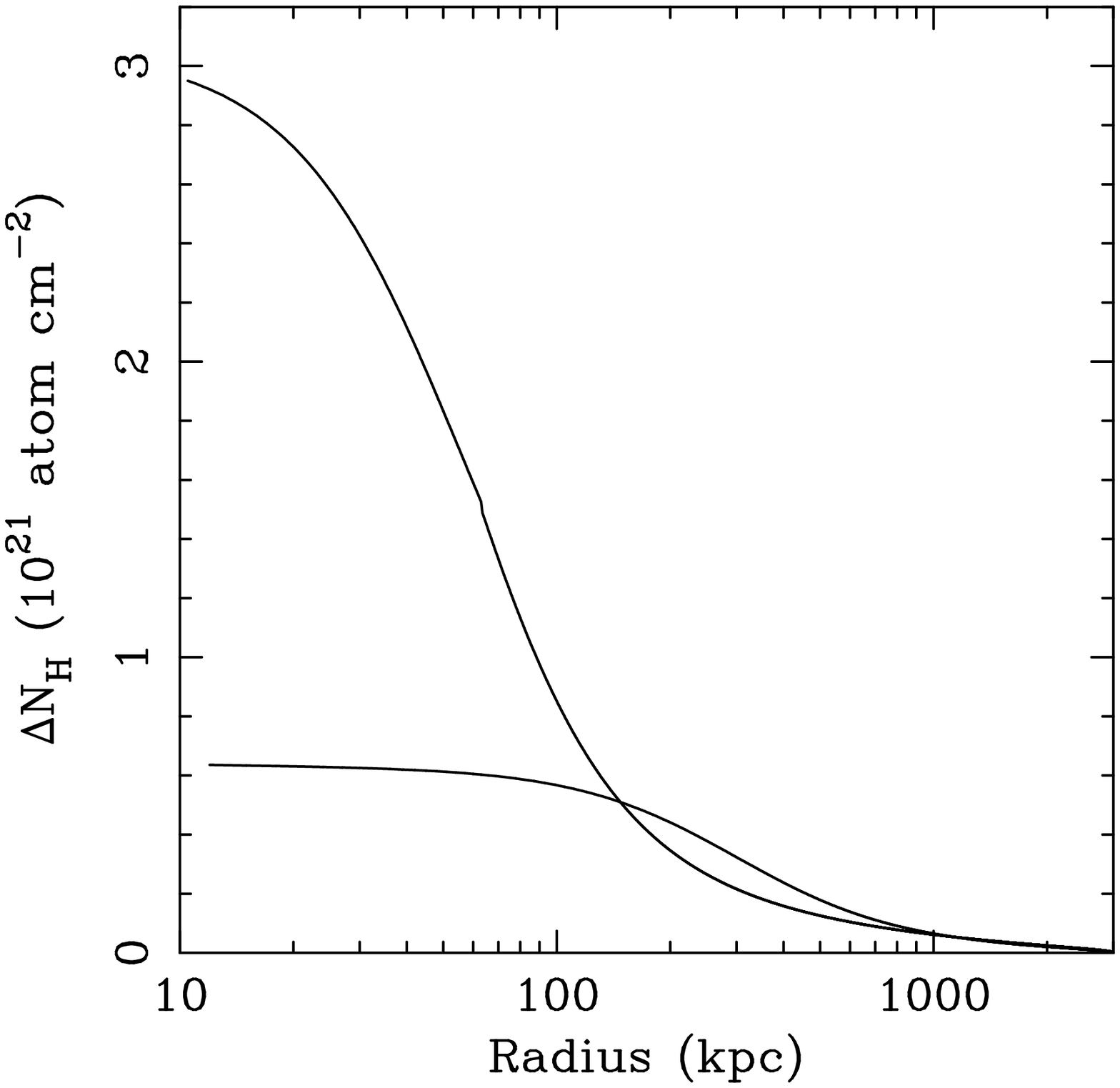,width=0.65\textwidth,angle=270}
}
\caption{(a) The equivalent column densities of hot hydrogen ions through 
two simulated clusters, one CF and one NCF system. The gas 
distributions have been modeled as isothermal spheres, with core radii of 
50 kpc (CF; upper curve) and 300 kpc (NCF; lower curve) respectively, and 
have been normalized to provide total X-ray gas masses within radii of 3Mpc 
of $2\times10^{14}$\Msun. The distribution of X-ray emitting gas 
is much more centrally concentrated in the CF cluster, particularly within the 
central $100-200$ kpc. (b) The X-ray absorbing column density, as a function of 
radius, due to dust distributed throughout the simulated CF 
(upper curve) and NCF (lower curve) clusters. The dust is assumed to 
have a multilayer distribution within the X-ray emitting gas, but be 
modeled as a simple uniform screen in front of the clusters.}
\end{figure*}

A final possibility, first discussed in detail by Voit \& Donahue (1995; see 
also Arnaud \& Mushotzky 1998), is that the X-ray absorption could be due 
to dust, with little or no associated gas. The K-absorption edge 
of oxygen is a primary source of X-ray absorption and its signature is 
fairly insensitive to the physical state of the oxygen, so long as 
it is not highly ionized. The introduction of a simple OIK absorption edge at 
$E \sim 0.54$ keV in the ASCA analysis (such as might be associated with 
oxygen-rich, silicate dust grains) typically provides at least as good a 
model for the intrinsic absorption as a cold gaseous absorber.
Arnaud \& Mushotzky (1998) have shown that $0.35-7.0$ keV Broad Band X-ray 
Telescope data for the Perseus cluster require excess absorption which is 
significantly better-explained by  a simple oxygen 
absorption edge (which they associate with dust) than by a cold, gaseous 
absorption model. Verification of this important result will be possible in 
the near future using observations made with the Chandra Observatory and XMM. 

  Optical and UV spectroscopy of 
the central emission-line nebulosities in cooling flows indicate 
the presence of significant amounts of dust (Hu 1992; Allen \etal 1995;
Crawford \etal 1999). Allen \etal (1995) also show that the column densities 
of absorbing gas inferred from the optical reddening studies are in reasonable 
agreement with the X-ray values, for Galactic dust/gas ratios. 
The infrared observations reported here and elsewhere, together with 
the sub-mm results of Edge \etal (1999), require the presence of 
significant dust masses in the core regions of at least a few nearby 
and/or exceptionally massive cooling flows (for which the best data exist) 
and are consistent with dust being a common feature of CF clusters. 
Limits on background counts of galaxies and
quasars behind clusters are consistent with extinctions ranging from
$A_{\rm B} = 0.2-0.5$ (\eg Boyle, Fong \& Shanks 1988; Romani \& Maoz
1992), although measurements of galaxy colours in nearby clusters 
constrain the reddening on cluster-wide scales to be $E(B-V)<0.06$ mag 
(Ferguson 1993). 

Voit \& Donahue (1995) argue that the levels of dust required 
to account for the observed X-ray absorption are unlikely to 
lie entirely within 
the cores of the clusters, and should be distributed more widely 
throughout the cluster gas. 
ROSAT results on the spatial distribution of the X-ray absorbing material 
(Allen \& Fabian 1997) show it to be centrally concentrated 
within the cooling radii of clusters and possibly be confined within 
cooling flows. However, the X-ray data are also consistent with 
a scenario in which the absorbing material is distributed in a manner 
similar to the X-ray emitting gas. This is illustrated in Fig. 8(a), where we 
show the equivalent column densities of hydrogen ions in the X-ray emitting 
gas through two simulated clusters; one CF and one NCF system. The gas 
distributions have been modeled as isothermal spheres, with core radii of 
50 and 300 kpc, respectively (and have been normalized to provide total 
X-ray gas masses within 3Mpc of the cluster centres of $2\times10^{14}$\Msun). 
We see that the distribution of X-ray emitting gas is much more centrally 
concentrated in the CF system, particularly within the central $\sim 100$ kpc. 

Fig. 8(b) shows the corresponding X-ray absorbing column densities 
as a function of radius (as would be determined from observations with 
ASCA-like CCD detectors in the $0.6-10.0$ keV band) in the 
case where the intracluster gas has an equivalent metallicity in dust grains 
of $\sim 0.2$ solar. For simplicity, we assume that the dust has an 
absorption spectrum in the
ASCA band similar to that of metal-rich cold gas (Morrison \& McCammon 1983; 
qualitatively similar results are obtained for an oxygen edge at 
0.54 keV). We also assume that the 
absorber follows a multilayer distribution within the X-ray emitting 
medium, but is modeled as a uniform absorbing screen in front of the emitting 
regions. 
The results imply mean emission-weighted column densities (where we assume the 
X-ray emissivity to be proportional to square of the gas density) of 
$\sim 2 \times 10^{21}$ \apc~for the CF cluster, and a few $\times 
10^{20}$ \apc~for the NCF system. (The absolute results on the 
excess column densities will depend upon the chemical composition of the 
dust). We thus see that if the X-ray 
absorption is due to dust, with a spatial distribution that follows the X-ray 
gas, then we can expect absorption signatures to be present in both CF and NCF 
clusters, but to be stronger in CF systems (assuming in each case that an 
appropriate spectral model is used in the analysis).
We would also expect the absorption to be concentrated 
towards the cores of CF clusters, in agreement with the observations.  
We note that if the X-ray absorption were entirely due to a pre-existing 
distribution of dust grains distributed throughout the
clusters, then the agreement between the observed masses of absorbing matter 
(calculated assuming that the absorption is due to cold gas confined 
within the cooling flows) and the masses predicted to have been 
accumulated by the cooling 
flows over their lifetimes (Fig. 6) must be regarded as 
coincidence. However, it remains possible that both 
material accumulated by the cooling flows and a more extended 
distribution of large dust grains (presumably due to supernovae enrichment 
at early epochs) could contribute to the X-ray absorption. 

The lifetime of grains of radius $a\mu$m to sputtering in hot gas of 
density $n$ is $\sim 2\times 10^6 a/n\yr$ (Draine \& Salpeter 1979). 
Provided that individual grains exceed 10$\mu$m in radius, they should 
survive for a Hubble time or longer throughout NCF clusters and beyond the 
cooling radius in clusters with cooling flows. Within cooling flows, the 
gas density rises inward so the grains are increasingly sputtered, releasing 
the metals into the gas phase which thus becomes increasingly metal rich 
towards the cluster centre. This provides one possible explanation for the 
abundance gradients inferred to be present in most CF clusters (Allen \&
Fabian 1998b; see also Irwin \& Bregman 1999). However, $10\mu$m grains are 
optically thick 
to soft X-rays and so some distribution of grain sizes extending to 
smaller values is required if dust is responsible for 
the observed X-ray absorption (\eg Laor \& Draine 1993). We note that the 
sputtering of small ($a \approxlt 0.1\mu$m) grains will significantly modify 
the reddening law in clusters with respect to the standard Galactic relation, 
reducing the optical/UV reddening (Laor \& Draine 1993).

In conclusion, the available data from other wavebands provide 
some support for the large column densities of intrinsic absorbing material 
inferred to be present from the X-ray data, at least in the innermost regions 
of cooling flows.  Optical and sub-mm observations of star formation in the 
cores of CF clusters and K-band observations of emission lines from molecular 
hydrogen suggest that significant masses of dusty, molecular gas are present 
in the central few tens of kpc of many CF clusters. The constraints from 21cm 
observations suggest that the bulk of
the material accumulated by cooling flows cannot remain in a long-lived 
reservoir of atomic hydrogen. Substantial masses of molecular gas may be 
distributed throughout cooling flows and as yet have avoided detection, but 
this material must be very cold ($T \sim 3$K) and dusty, a possibility that 
remains controversial. 
The very large X-ray column densities ($\approxgt 10^{22}$ \apc) 
required if the absorbing material has a multilayer distribution 
are probably incompatible with a gaseous absorber. 
It remains possible that dust grains, either present in the cooled 
material deposited by the cooling flows or distributed 
throughout the clusters in a manner similar to the X-ray gas, 
may be responsible for much of the observed X-ray absorption.

\section{Conclusions}

The main conclusions that may be drawn from this paper may be summarized as 
follows:
\vskip 0.2cm

 (i) We have demonstrated the need for multiphase models to consistently
explain the spectral and imaging X-ray data for the CF clusters included in
our study. The mass deposition rates from the cooling flows, 
independently inferred from multiphase analyses of the ASCA spectra and 
deprojection analyses of the ROSAT HRI images, exhibit good agreement,
especially once the effects of intrinsic X-ray absorption have been 
accounted for in a consistent manner. The mass deposition rates from the
largest cooling flows exceed 1000 \Msunpyr, identifying these as
some of the most massive cooling flows known.  

 (ii) We have confirmed the presence of intrinsic X-ray absorption in 
the cluster spectra using a variety of spectral models. The measured 
equivalent hydrogen column densities are sensitive to the spectral models used 
in the analysis, ranging from a few $10^{20}$ \apc~for a 
simple isothermal emission model to a few $10^{21}$ \apc~using 
our preferred cooling-flow models, assuming in each case that the
absorber lies in a uniform foreground screen. Both the CF and NCF systems 
exhibit excess absorption at a similar level (on average) 
when analysed with the same, simple isothermal model. 

  (iii) The masses of X-ray absorbing material inferred to be present in
the CF clusters (assuming the absorption to be due to cold gas
with a covering fraction of unity) are in 
reasonable agreement with the masses expected to have been accumulated 
by the cooling flows over their lifetimes. 

  (iv) The ASCA spectra constrain the covering fraction of 
the absorbing material acting on the cooling flows to be close to (or exceed) 
unity. If the X-ray absorption is due to many small, similarly-sized 
clouds along each line of sight, intermixed with the X-ray emitting gas, 
then the column densities inferred from 
the spectral analysis may significantly underestimate the true column 
densities of absorbing material in the cooling flows.

   (vii) We have summarized the constraints on the physical properties of
the X-ray absorbing material from observations in other wavebands. 
Substantial 60 and 100$\mu$m fluxes, sufficient to account for the bolometric 
luminosities absorbed and reprocessed within the clusters, are 
detected from several of the largest cooling flows. Optical observations of 
star formation in the cores of cooling flows also suggests that significant 
masses of molecular gas are present in the central few tens of kpc of 
many CF systems. Constraints from 21cm observations suggest that the 
bulk of the material accumulated within the cooling radii cannot remain in a 
long-lived  reservoir of atomic hydrogen. Substantial masses of molecular 
gas may be distributed throughout cooling flows and as yet have avoided 
detection, although this material must be very cold ($T \sim 3$K), a 
possibility that remains controversial. The very large X-ray column 
densities ($\approxgt 10^{22}$ \apc) required if the absorbing material has 
a multilayer distribution are probably incompatible with a 
gaseous absorber. It remains possible that dust grains, either present 
in the material accumulated by the cooling flows or distributed more widely 
in a manner similar to the X-ray gas, may be responsible for much of the 
X-ray absorption.

\section*{Acknowledgments}

I thank Andy Fabian, Roderick Johnstone and Dave White for  
many helpful discussions and Harald Ebeling, Alastair Edge and 
Carolin Crawford for their continuing efforts with the BCS project. I 
acknowledge the support of the Royal Society.


\clearpage

\begin{table*}
\vskip 0.2truein
\begin{center}
\caption{Summary of the ASCA Observations}
\vskip 0.2truein
\begin{tabular}{ c c c c c c c c c c c c }
\multicolumn{1}{c}{} &
\multicolumn{1}{c}{} &
\multicolumn{1}{c}{} &
\multicolumn{1}{c}{} &
\multicolumn{1}{c}{} &
\multicolumn{1}{c}{} &
\multicolumn{1}{c}{} &
\multicolumn{4}{c}{} &
\multicolumn{1}{l}{} \\
 \hline                                                                               
Cluster         & ~ &  z     & $N_{\rm H}$  & ~ &  Date     & ~ &    S0  &   S1  &  G2   &  G3  \\  
 \hline                                                                               
&&&&&&&&&&& \\                                                                         
Abell 2744      & ~ & 0.308  & 0.16      & ~ &  1994 Jul 04  & ~ &  37605 & 26086 & 62749 & 62753   \\       
Abell 478$^\dagger$  & ~ & 0.088  & 3.00     & ~ &  1994 Feb 24  & ~ & 27953 & ----- & 34560 & 34554  \\        
Abell 520       & ~ & 0.203  & 0.78      & ~ &  1994 Sep 09  & ~ &  16802 & 15871 & 17656 & 17617 \\    
Abell 586       & ~ & 0.171  & 0.52      & ~ &  1994 Mar 22  & ~ &  8482  & ----- & 17668 & 17648 \\      
PKS0745-191     & ~ & 0.103  & 4.24     & ~ &  1993 Nov 06  & ~ &  29146 & ----- & 37553 & 37553 \\      
Abell 665       & ~ & 0.182  & 0.42      & ~ &  1993 Sep 18  & ~ &  22926 & 19869 & 34815 & 34799 \\      
IRAS 09104+4109 & ~ & 0.442  & 0.10      & ~ &  1993 Nov 12  & ~ &  33447 & 24348 & 38655 & 38655 \\       
Abell 773       & ~ & 0.217  & 0.14      & ~ &  1994 Nov 22  & ~ &  38500 & 38086 & 40841 & 40831 \\    
Abell 963       & ~ & 0.206  & 0.14      & ~ &  1993 Apr 22  & ~ &  29611 & 29039 & 29883 & 29881 \\       
Zwicky 3146     & ~ & 0.291  & 0.30      & ~ &  1993 May 18  & ~ &  29663 & 28643 & 32622 & 32632 \\      
Abell 1068      & ~ & 0.139  & 0.14      & ~ &  1996 Dec 06  & ~ &  23777 & 23645 & 21463 & 21639 \\       
Abell 1413      & ~ & 0.143  & 0.22      & ~ &  1993 Dec 11  & ~ &  26084 & 19862 & 36173 & 36175 \\  
Abell 1689      & ~ & 0.184  & 0.18      & ~ &  1993 Jun 26  & ~ &  29575 & 23642 & 37817 & 37817 \\  
Abell 1704      & ~ & 0.216  & 0.18      & ~ &  1994 Apr 16  & ~ &  18847 & 11784 & 20598 & 20596 \\  
RXJ1347.5-1145  & ~ & 0.451  & 0.49      & ~ &  1995 Jan 17  & ~ &  27882 & 17549 & 38968 & 38958 \\      
Abell 1795      & ~ & 0.063  & 0.12      & ~ &  1993 Jun 16  & ~ &  31284 & ----- & 37649 & 37641 \\ 
MS1358.4+6245   & ~ & 0.327  & 0.19      & ~ &  1995 Apr 27  & ~ &  32532 & 30815 & 31981 & 31513 \\  
Abell 1835 \#1  & ~ & 0.252  & 0.23      & ~ &  1994 Jul 20  & ~ &  18051 & 17532 & 17464 & 17460 \\       
Abell 1835 \#2  & ~ & $""$   & $""$     & ~ &  1994 Jul 21  & ~ &  16876 & 16444 & 16412 & 16410 \\       
MS1455.0+2232   & ~ & 0.258  & 0.32      & ~ &  1994 Jul 18  & ~ &  29388 & 28539 & 28903 & 28895 \\       
Abell 2029      & ~ & 0.077  & 0.31      & ~ &  1994 Feb 19  & ~ &  34211 & 32420 & 35038 & 34963 \\      
Abell 2142      & ~ & 0.089  & 0.42      & ~ &  1994 Feb 21  & ~ &  12335 & ----- & 15768 & 15766 \\      
Abell 2163      & ~ & 0.208  & 1.21     & ~ &  1993 Aug 08  & ~ &  25126 & 18224 & 32760 & 32322 \\       
Abell 2204      & ~ & 0.152  & 0.57      & ~ &  1994 Aug 20  & ~ &  12749 & 11865 & 14698 & 14698 \\     
Abell 2218      & ~ & 0.175  & 0.32      & ~ &  1993 Apr 30  & ~ &  28241 & 26054 & 37970 & 37968 \\       
Abell 2219      & ~ & 0.228  & 0.18      & ~ &  1994 Aug 07  & ~ &  32705 & 31697 & 35849 & 35849 \\      
Abell 2261      & ~ & 0.224  & 0.33      & ~ &  1996 Feb 24  & ~ &  19224 & 18823 & 19249 & 19219 \\      
Abell 2319      & ~ & 0.056  & 0.80      & ~ &  1993 Jul 21  & ~ &  9842  & 5007  & 13169 & 13404 \\      
MS2137.3-2353   & ~ & 0.313  & 0.36      & ~ &  1994 May 08  & ~ &  15167 & 15732 & 17035 & 17056 \\      
Abell 2390      & ~ & 0.233  & 0.68      & ~ &  1994 Nov 13  & ~ &  6172  & 2632  & 10340 & 10338 \\      
AC114           & ~ & 0.312  & 0.13      & ~ &  1995 Nov 09  & ~ &  36739 & 36295 & 35987 & 35971 \\      

&&&&&&&&&&& \\                                                                         
\hline 
&&&&&&&&&&& \\                                                                         
\end{tabular}
\end{center}
\parbox {7in}
{Notes: Columns 2 and 3 list the redshift and Galactic column density, in 
units of $10^{21}$ \apc~from Dickey \& Lockman (1990), with the exception
of Abell 478$^{\dagger}$, for which the value is from Allen \& Fabian (1997).  
For PKS0745-191, this study (Section 3.2) suggests a Galactic column density 
of $\sim 3.5 \times 10^{21}$ \apc, which is lower than the Dickey \& Lockman 
(1990) value. Column 4 lists the dates of the ASCA observation. Columns $5-8$ 
list the net exposure times (in seconds) in each of the four ASCA detectors, 
after all screening and cleaning procedures were carried out. }
\end{table*}

\clearpage

\begin{table}
\vskip 0.2truein
\begin{center}
\caption{Extraction radii for the ASCA spectra}
\vskip 0.2truein
\begin{tabular}{ c c c c c }
\multicolumn{1}{c}{} &
\multicolumn{1}{c}{} &
\multicolumn{1}{c}{} &
\multicolumn{1}{c}{} &
\multicolumn{1}{c}{} \\
\hline                                                                               
Cluster         & ~ & S0 & S1     &  Chip mode \\  
                & ~ & (amin/kpc) & (amin/kpc)   &           \\
\hline                                                                           
&&&& \\                                                                     
Abell 2744      & ~ & 3.4/1140  & 2.7/905       & 4(1)    \\       
Abell 478       & ~ & 3.8/503   & ---           & 4(1)    \\        
Abell 520       & ~ & 3.6/922   & 3.1/794       & 2(2)    \\    
Abell 586       & ~ & 3.6/814   & ---           & 4(1)    \\      
PKS0745-191     & ~ & 4.2/636   & ---           & 4(1)    \\      
Abell 665       & ~ & 4.5/1070  & 5.3/1260      & 4(2)    \\      
IRAS 09104+4109 & ~ & 4.0/1620  & 4.0/1620      & 4(4)    \\       
Abell 773       & ~ & 4.4/1180  & 3.9/1050      & 1       \\    
Abell 963       & ~ & 3.6/932   & 2.8/725       & 2(1)    \\       
Zwicky 3146     & ~ & 5.3/1720  & 4.3/1390      & 2(2)    \\      
Abell 1068      & ~ & 3.4/656   & 2.6/502       & 1       \\       
Abell 1413      & ~ & 4.0/790   & 4.0/790       & 4(4)    \\  
Abell 1689      & ~ & 4.0/955   & 4.0/955       & 4(4)    \\  
Abell 1704      & ~ & 3.4/909   & 2.6/695       & 4(1)    \\  
RXJ1347.5-1145  & ~ & 4.0/1630  & 4.0/1630      & 4(1)    \\      
Abell 1795      & ~ & 4.5/445   & ---           & 4(2)    \\ 
MS1358.4+6245   & ~ & 3.9/1350  & 2.9/1010      & 4(2)    \\  
Abell 1835 \#1  & ~ & 4.1/1220  & 3.2/948       & 2(1)    \\       
Abell 1835 \#2  & ~ & 4.1/1220  & 3.2/948       & 2(1)    \\       
MS1455.0+2232   & ~ & 3.7/1110  & 2.8/842       & 2(1)    \\       
Abell 2029      & ~ & 3.5/413   & 2.7/319       & 2(1)    \\      
Abell 2142      & ~ & 6.0/802   & ---           & 4(4)    \\      
Abell 2163      & ~ & 3.7/964   & 4.5/1170      & 4(2)    \\       
Abell 2204      & ~ & 4.9/1010  & 4.7/973       & 2(1)    \\     
Abell 2218      & ~ & 3.7/852   & 2.5/575       & 4(1)    \\       
Abell 2219      & ~ & 5.3/1470  & 4.5/1250      & 2(1)    \\      
Abell 2261      & ~ & 4.5/1230  & 3.6/987       & 2(2)    \\      
Abell 2319      & ~ & 4.2/373   & 4.0/355       & 4(1)    \\      
MS2137.3-2353   & ~ & 4.3/1460  & 3.4/1150      & 2(2)    \\      
Abell 2390      & ~ & 3.6/1010  & 2.5/704       & 4(1)    \\      
AC114           & ~ & 3.4/1150  & 2.9/980       & 1       \\      
&&&& \\                                                                     
\hline 
&&&& \\                                                                     
\end{tabular}
\end{center}
\parbox {3.5in}
{The radii of the circular extraction regions for the ASCA SIS data (in
arcmin and kpc) and the chip modes used in the observations (either 1,2
or 4-CCD mode). The numbers in parentheses indicate the number of
chips contributing to the extracted spectra. For the GIS data a fixed
extraction radius of 6 arcmin was used. For Abell 2142, the 2 arcmin (3
arcmin) radius region surrounding the X-ray bright Seyfert-1 galaxy 1556+274 
was masked out and excluded from the analysis for the SIS (GIS) data. 
}
\end{table}

\clearpage

\begin{table*}
\vskip 0.2truein
\begin{center}
\caption{Summary of the ROSAT Observations}
\vskip 0.2truein
\begin{tabular}{ c c c c c c c }
\multicolumn{1}{c}{} &
\multicolumn{1}{c}{} &
\multicolumn{1}{c}{} &
\multicolumn{1}{c}{} &
\multicolumn{1}{c}{} &
\multicolumn{2}{c}{} \\                            
\hline                                                                                                                               
 Cluster     & ~ &  Date     & Exposure (s)    & ~ &     R.A.  (J2000). &    Dec. (J2000.)                 \\  
\hline                                                                                                                               
&&&&&& \\                                                                                                                         
Abell 2744         & ~ &  1994 Dec 09  & 34256   & ~ &   $00^{\rm h}14^{\rm m}18.7^{\rm s}$ & $-30^{\circ}23'11''$  \\  
Abell 478          & ~ &  1991 Feb 10  & 22840   & ~ &   $04^{\rm h}13^{\rm m}25.4^{\rm s}$ & $10^{\circ}27'58''$  \\  
Abell 520          & ~ &  1994 Mar 05  & 12728   & ~ &   $04^{\rm h}54^{\rm m}10.1^{\rm s}$ &  $02^{\circ}55'27''$  \\  
Abell 586          & ~ &  1994 Oct 06  & 23136   & ~ &   $07^{\rm h}32^{\rm m}20.4^{\rm s}$ & $31^{\circ}37'55''$  \\  
PKS0745-191        & ~ &  1992 Oct 20  & 23750   & ~ &   $07^{\rm h}47^{\rm m}31.1^{\rm s}$ & $-19^{\circ}17'47''$  \\  
Abell 665(P)       & ~ &  1991 Apr 10  & 38308   & ~ &   $08^{\rm h}30^{\rm m}58.9^{\rm s}$ & $65^{\circ}50'37''$  \\  
IRAS09104+4109 \#1 & ~ &  1994 Nov 08  & 7968    & ~ &   $09^{\rm h}13^{\rm m}45.5^{\rm s}$ & $40^{\circ}56'29''$  \\  
IRAS09104+4109 \#2 & ~ &  1995 Apr 12  & 21904   & ~ &   "" & "" \\  
Abell 773          & ~ &  1995 Apr 15  & 6240    & ~ &   $09^{\rm h}17^{\rm m}53.4^{\rm s}$ &  $51^{\circ}43'29''$  \\  
Abell 963          & ~ &  1992 Nov 24  & 10104   & ~ &   $10^{\rm h}17^{\rm m}03.4^{\rm s}$ & $39^{\circ}02'51''$  \\  
Zwicky 3146 \#1    & ~ &  1992 Nov 27  & 15214   & ~ &   $10^{\rm h}23^{\rm m}39.8^{\rm s}$ & $04^{\circ}11'11''$  \\
Zwicky 3146 \#2    & ~ &  1993 May 17  & 10831   & ~ &   "" & "" \\
Abell 1068         & ~ &  1995 Apr 29  & 15760   & ~ &   $10^{\rm h}40^{\rm m}44.6^{\rm s}$ & $39^{\circ}57'12''$  \\  
Abell 1413(P)      & ~ &  1991 Nov 27  & 7696    & ~ &   $11^{\rm h}55^{\rm m}18.7^{\rm s}$ & $23^{\circ}24'12''$  \\  
Abell 1689 \#1     & ~ &  1994 Jul 22  & 13080   & ~ &   $13^{\rm h}11^{\rm m}29.1^{\rm s}$ & $-01^{\circ}20'40''$  \\  
Abell 1689 \#2     & ~ &  1995 Jun 24  & 9648    & ~ &   "" & "" \\  
Abell 1704         & ~ &  1995 Apr 17  & 41360   & ~ &   $13^{\rm h}14^{\rm m}24.8^{\rm s}$ & $64^{\circ}34'39''$  \\  
RXJ1347.5-1145     & ~ &  1995 Jan 28  & 15760   & ~ &   $13^{\rm h}47^{\rm m}31^{\rm s}$ & $-11^{\circ}45'11''$ \\
Abell 1795 \#1     & ~ &  1992 Jun 25  & 2768    & ~ &   $13^{\rm h}48^{\rm m}52.7^{\rm s}$ & $26^{\circ}35'27''$  \\  
Abell 1795 \#2     & ~ &  1993 Jan 21  & 11088   & ~ &   "" & "" \\
Abell 1795 \#3     & ~ &  1994 Jun 23  & 11080   & ~ &   "" & "" \\  
MS1358.4+6245 \#1  & ~ &  1991 Nov 05  & 2528    & ~ &   $13^{\rm h}59^{\rm m}51.0^{\rm s}$ & $62^{\circ}31'04''$  \\  
MS1358.4+6245 \#2  & ~ &  1993 May 14  & 15872   & ~ &   "" & "" \\  
Abell 1835         & ~ &  1993 Jan 22  & 2850    & ~ &   $14^{\rm h}01^{\rm m}02.0^{\rm s}$ & $02^{\circ}52'40''$  \\ 
MS1455.0+2232 \#1  & ~ &  1992 Jan 11  & 4088    & ~ &   $14^{\rm h}57^{\rm m}15.0^{\rm s}$ & $22^{\circ}20'36''$  \\
MS1455.0+2232 \#2  & ~ &  1993 Jan 20  & 4230    & ~ &   "" & "" \\
MS1455.0+2232 \#3  & ~ &  1994 Jul 07  & 6584    & ~ &   "" & "" \\
Abell 2029         & ~ &  1990 Jul 25  & 17758   & ~ &   $15^{\rm h}10^{\rm m}56.2^{\rm s}$ & $05^{\circ}44'42''$  \\  
Abell 2142         & ~ &  1994 Jul 26  & 19776   & ~ &   $15^{\rm h}58^{\rm m}20.1^{\rm s}$ & $27^{\circ}13'52''$  \\  
Abell 2163         & ~ &  1994 Aug 13  & 36248   & ~ &   $16^{\rm h}15^{\rm m}45.9^{\rm s}$ & $-06^{\circ}08'58''$  \\  
Abell 2204         & ~ &  1995 Sep 01  & 15472   & ~ &   $16^{\rm h}32^{\rm m}47.0^{\rm s}$ & $05^{\circ}34'33''$  \\  
Abell 2218 \#1     & ~ &  1994 Jan 05  & 11520   & ~ &   $16^{\rm h}35^{\rm m}52.5^{\rm s}$ & $66^{\circ}12'29''$  \\  
Abell 2218 \#2     & ~ &  1994 Apr 05  & 27016   & ~ &   "" & "" \\  
Abell 2218 \#3     & ~ &  1994 Jun 14  & 30016   & ~ &   "" & "" \\  
Abell 2218 \#4     & ~ &  1994 Jun 17  & 24304   & ~ &   "" & "" \\  
Abell 2219         & ~ &  1994 Jan 17  & 13242   & ~ &   $16^{\rm h}40^{\rm m}20.2^{\rm s}$ & $46^{\circ}42'29''$  \\  
Abell 2261         & ~ &  1995 Feb 11  & 16288   & ~ &   $17^{\rm h}22^{\rm m}27.3^{\rm s}$ & $32^{\circ}07'58''$  \\  
Abell 2319         & ~ &  1991 Mar 22  & 5444    & ~ &   $19^{\rm h}21^{\rm m}12.2^{\rm s}$ & $43^{\circ}56'38''$  \\  
MS2137.3-2353      & ~ &  1994 Apr 24  & 13656   & ~ &   $21^{\rm h}40^{\rm m}15.2^{\rm s}$ & $-23^{\circ}39'41''$  \\  
Abell 2390         & ~ &  1993 Nov 23  & 27764   & ~ &   $21^{\rm h}53^{\rm m}36.5^{\rm s}$ & $17^{\circ}41'45''$  \\  
AC114 \#1          & ~ &  1993 May 17  & 10048   & ~ &   $22^{\rm h}58^{\rm m}48.7^{\rm s}$ & $-34^{\circ}48'19''$  \\  
AC114 \#2          & ~ &  1994 May 09  & 13144   & ~ &   "" & "" \\  

&&&&&& \\                                                                                                                         
\hline
&&&&&& \\                                                                                                                         
\end{tabular}
\end{center}
\parbox {7in}
{Notes: Columns 2 and 3 list the dates of observation and exposure times (in seconds).
Where more than a single observation of a cluster was made, details 
for each observation are listed. 
Columns 4 and 5 give the coordinates of the centroids of the X-ray
emission from the clusters. For Abell 665 and 1413, HRI images were not 
available at the time of writing and PSPC data were used for the imaging
analysis. This is indicated by a (P) after the cluster name. 
}
\end{table*}

\clearpage

\begin{table*}
\vskip 0.2truein
\begin{center}
\caption{Results from the spectral analysis} 
\vskip 0.2truein
\begin{tabular}{ c c c c c c c c c c c }
&&&&&&&&&&  \\                                                                                                                                               
\hline                                                                                                                                                         
            & ~ &   Parameters   & ~ &     Model A            & ~~~ & Model B                 & ~~~ &     Model C    & ~~~ &  Model D     \\
\hline                                                                                                                                                         
&&&&&&&&&&  \\                                                                                                                                               
            & ~ &   $kT$         & ~ & $11.04^{+0.79}_{-0.73}$ & ~~~ & $7.75^{+0.59}_{-0.53}$ & ~~~ &  ---   & ~~~ & $8.5^{+0.9}_{-0.7}$ \\ 
            & ~ &   $Z$          & ~ & $0.17^{+0.09}_{-0.09}$ & ~~~ & $0.16^{+0.06}_{-0.06}$  & ~~~ &  ---   & ~~~ & $0.17^{+0.08}_{-0.07}$ \\ 
Abell 2744  & ~ &   $N_{\rm H}$  & ~ & $0.16$                 & ~~~ & $1.44^{+0.21}_{-0.21}$  & ~~~ &  ---   & ~~~ & $1.25^{+0.14}_{-0.22}$  \\ 
            & ~ &   ${\dot M_{\rm S}}$   & ~ &  ---                   & ~~~ &  ---                    & ~~~ &  ---   & ~~~ & ---  \\ 
            & ~ &   $kT_2$       & ~ &  ---                   & ~~~ &  ---                    & ~~~ &  ---   & ~~~ & $<1.7$ \\
\end{tabular}
\end{center}
 
\parbox {7in}
{}
\end{table*}

\clearpage

\addtocounter{table}{-1}
\begin{table*}
\vskip 0.2truein
\begin{center}
\caption{Spectral Results - continued}
\vskip 0.2truein
\begin{tabular}{ c c c c c c c c c c c }
&&&&&&&&&&  \\                                                                                                                                               
\hline
            & ~ &   Parameters   & ~ &     Model A            & ~~~ & Model B                 & ~~~ &     Model C    & ~~~ &  Model D     \\
\hline
&&&&&&&&&&  \\                                                                                                                                               
            & ~ &   $kT$         & ~ & $9.03^{+0.58}_{-0.52}$ & ~~~ & $8.12^{+0.62}_{-0.54}$ & ~~~ &  ---  & ~~~ & $14.2^{+10.4}_{-5.2}$ \\           
            & ~ &   $Z$          & ~ & $0.22^{+0.07}_{-0.08}$ & ~~~ & $0.23^{+0.06}_{-0.07}$ & ~~~ &  ---  & ~~~ & $0.33^{+0.11}_{-0.13}$ \\           
Abell 665   & ~ &   $N_{\rm H}$  & ~ & $0.42 $                & ~~~ & $0.72^{+0.14}_{-0.14}$ & ~~~ &  ---  & ~~~ & $0.0^{+0.6}_{-0.0}$ \\           
            & ~ &   ${\dot M_{\rm S}}$   & ~ &  ---           & ~~~ &        ---             & ~~~ &  ---  & ~~~ & ---  \\           
            & ~ &   $kT_2$       & ~ &  ---                   & ~~~ &  ---                   & ~~~ &  ---  & ~~~ & $<4.4$ \\
\end{tabular}
\end{center}
 
\parbox {7in}
{}
\end{table*}

\clearpage

\addtocounter{table}{-1}
\begin{table*}
\vskip 0.2truein
\begin{center}
\caption{Spectral Results - continued}
\vskip 0.2truein
\begin{tabular}{ c c c c c c c c c c c }
&&&&&&&&&&  \\                                                                                                                                               
\hline                                                                                                                                                         
            & ~ &   Parameters   & ~ &     Model A            & ~~~ & Model B                & ~~~ &     Model C        & ~~~ &  Model D     \\
\hline                                                                                                                                                         
&&&&&&&&&&  \\                                                                                                                                               
            & ~ &   $kT$         & ~ & $4.21^{+0.18}_{-0.14}$ & ~~~ & $3.46^{+0.16}_{-0.13}$ & ~~~ & $5.5^{+1.4}_{-0.9}$    & ~~~ & $16.6^{+\infty}_{-8.6}$ \\    
            & ~ &   $Z$          & ~ & $0.43^{+0.08}_{-0.08}$ & ~~~ & $0.49^{+0.08}_{-0.08}$ & ~~~ & $0.42^{+0.10}_{-0.08}$ & ~~~ & $0.53^{+0.10}_{-0.11}$ \\     
Abell 1068  & ~ &   $N_{\rm H}$  & ~ & $0.14$                 & ~~~ & $0.99^{+0.15}_{-0.14}$ & ~~~ & $0.14$                 & ~~~ & $0.14$                 \\ 
            & ~ &   ${\dot M_{\rm S}}$   & ~ &  ---           & ~~~ &  ---                   & ~~~ & $1345^{+145}_{-177}$   & ~~~ & --- \\       
            & ~ &   $kT_2$       & ~ &  ---                   & ~~~ &  ---                   & ~~~ & ---                    & ~~~ & $2.5^{+0.3}_{-0.6}$ \\
\end{tabular}
\end{center}
 
\parbox {7in}
{}
\end{table*}

\clearpage

\addtocounter{table}{-1}
\begin{table*}
\vskip 0.2truein
\begin{center}
\caption{Spectral Results - continued}
\vskip 0.2truein
\begin{tabular}{ c c c c c c c c c c c }
&&&&&&&&&&  \\                                                                                                                                               
\hline                                                                                                                                                         
            & ~ &   Parameters   & ~ &     Model A            & ~~~ & Model B                & ~~~ &     Model C        & ~~~ &  Model D     \\
\hline                                                                                                                                                         
&&&&&&&&&&  \\                                                                                                                                               
            & ~ &   $kT$         & ~ & $5.40^{+0.08}_{-0.09}$ & ~~~ & $5.33^{+0.10}_{-0.11}$ & ~~~ & $5.87^{+0.29}_{-0.25}$ & ~~~ & $7.1^{+4.6}_{-0.7}$ \\    
            & ~ &   $Z$          & ~ & $0.36^{+0.03}_{-0.02}$ & ~~~ & $0.36^{+0.03}_{-0.02}$ & ~~~ & $0.36^{+0.03}_{-0.02}$ & ~~~ & $0.38^{+0.04}_{-0.04}$ \\     
Abell 1795  & ~ &   $N_{\rm H}$  & ~ & $0.12$                 & ~~~ & $0.17^{+0.05}_{-0.05}$ & ~~~ & $0.12$                 & ~~~ & $0.12$                 \\
            & ~ &   ${\dot M_{\rm S}}$   & ~ &  ---           & ~~~ &  ---                   & ~~~ & $300^{+104}_{-110}$    & ~~~ & --- \\       
            & ~ &   $kT_2$       & ~ &  ---                   & ~~~ &  ---                   & ~~~ & ---                    & ~~~ & $2.5^{+1.3}_{-0.6}$ \\
\end{tabular}
\end{center}
 
\parbox {7in}
{}
\end{table*}

\clearpage

\addtocounter{table}{-1}
\begin{table*}
\vskip 0.2truein
\begin{center}
\caption{Spectral Results - continued}
\vskip 0.2truein
\begin{tabular}{ c c c c c c c c c c c }
&&&&&&&&&&  \\                                                                                                                                               
\hline                                                                                                                                                         
            & ~ &   Parameters   & ~ &     Model A            & ~~~ & Model B                & ~~~ &     Model C        & ~~~ &  Model D     \\
\hline                                                                                                                                                         
&&&&&&&&&&  \\                                                                                                                                               
            & ~ &   $kT$         & ~ & $8.67^{+0.32}_{-0.30}$  & ~~~ & $7.82^{+0.35}_{-0.32}$ & ~~~ &  $9.3^{+1.3}_{-0.7}$    & ~~~ & $10.0^{+1.5}_{-0.7}$  \\        
            & ~ &   $Z$          & ~ & $0.25^{+0.05}_{-0.05}$  & ~~~ & $0.25^{+0.04}_{-0.05}$ & ~~~ &  $0.27^{+0.05}_{-0.05}$ & ~~~ & $0.30^{+0.07}_{-0.07}$ \\ 
Abell 2142  & ~ &   $N_{\rm H}$  & ~ & $0.42$                  & ~~~ & $0.77^{+0.10}_{-0.10}$ & ~~~ &  $0.42$                 & ~~~ & $0.42$                  \\   
            & ~ &   ${\dot M_{\rm S}}$   & ~ &  ---            & ~~~ & ---                    & ~~~ &  $529^{+177}_{-164}$    & ~~~ & --- \\                    
            & ~ &   $kT_2$       & ~ &  ---                    & ~~~ & ---                    & ~~~ &  ---                    & ~~~ & $0.9^{+1.6}_{-0.3}$ \\
\end{tabular}
\end{center}
 
\parbox {7in}
{}
\end{table*}

\clearpage

\addtocounter{table}{-1}
\begin{table*}
\vskip 0.2truein
\begin{center}
\caption{Spectral Results - continued}
\vskip 0.2truein
\begin{tabular}{ c c c c c c c c c c c }
&&&&&&&&&&  \\                                                                                                                                               
\hline                                                                                                                                                         
            & ~ &   Parameters   & ~ &     Model A            & ~~~ & Model B                & ~~~ &     Model C        & ~~~ &  Model D     \\
\hline                                                                                                                                                         
&&&&&&&&&&  \\                                                                                                                                               
            & ~ &   $kT$         & ~ & $8.82^{+0.61}_{-0.54}$ & ~~~ & $6.64^{+0.51}_{-0.46}$  & ~~~ & $10.9^{+5.9}_{-2.2}$   & ~~~ & $13.4^{+6.5}_{-2.6}$ \\
            & ~ &   $Z$          & ~ & $0.32^{+0.10}_{-0.09}$ & ~~~ & $0.31^{+0.07}_{-0.08}$  & ~~~ & $0.37^{+0.09}_{-0.09}$ & ~~~ & $0.52^{+0.17}_{-0.15}$ \\  
Abell 2261  & ~ &   $N_{\rm H}$  & ~ & $0.33$                 & ~~~ & $1.17^{+0.17}_{-0.17}$  & ~~~ & $0.33$                 & ~~~ & $0.33$                 \\
            & ~ &   ${\dot M_{\rm S}}$   & ~ &  ---           & ~~~ &  ---                    & ~~~ & $1323^{+215}_{-263}$   & ~~~ & --- \\                
            & ~ &   $kT_2$       & ~ &  ---                   & ~~~ &  ---                    & ~~~ & ---                    & ~~~ & $2.5^{+1.2}_{-0.7}$ \\
\end{tabular}
\end{center}
 
\parbox {7in}
{}
\end{table*}

\addtocounter{table}{-1}
\begin{table*}
\vskip 0.2truein
\begin{center}
\caption{Spectral Results}
\begin{tabular}{ c c c c c c c c c c c }
\end{tabular}
\end{center}
 
\parbox {7in}
{The best-fit parameter values and 90 per cent  ($\Delta \chi^2 =
2.71$) confidence limits from the spectral analysis of the ASCA data. 
Temperatures ($kT$), metallicities ($Z$), column densities ($N_{\rm H}$) and
intrinsic column densities ($\Delta N_{\rm H}$) were linked to take the same
values in all four detectors. With spectral models C and D, the mass 
deposition rates of the cooling flows (${\dot M_{\rm S}}$) and luminosities of 
the cooler spectral components were also linked to take the same values, but 
scaled by a normalization factor proportional to the total flux measured in 
the detector (Section 3.1).  Only the normalization of the hotter isothermal 
emission component was allowed to vary independently for each detector. 
Temperatures are in keV, metallicities as a fraction of the solar value 
(Anders \& Grevesse 1989), column densities and intrinsic column densities in 
units of $10^{21}$ atom cm$^{-2}$ and mass deposition rates in \Msunpyr. 
The ${\dot M_{\rm S}}$~values for model C are measured with the G3 detector. 
$L_{\rm X}$ values are the $2-10$ keV X-ray luminosities in the rest
frame of the source. $L_{\rm Bol}$ values are the bolometric luminosities, corrected for 
the effects of Galactic and intrinsic absorption. All luminosities are in 
units of $10^{44}$\ergps~and are measured in the G3 detector. An entry of U.C. 
indicates that a  parameter was unconstrained by the data.
Note that the analysis of PKS0745-191 with spectral model C has the
Galactic column density included as a free parameter (Section 3.2).
The fit to Abell 2390 with spectral Model D has a second possible solution 
with $kT \sim 0.2$ keV and $kT_2 \sim 9.0$ keV which provides a comparable 
$\chi^2$, although this solution appears physically implausible.}
\end{table*}

\clearpage

\begin{table*}
\vskip 0.2truein
\begin{center}
\caption{The goodness of fit for the different spectral models}
\vskip 0.2truein
\begin{tabular}{ c c c c c c c c c c }
\hline                                                                               
\multicolumn{1}{c}{} &
\multicolumn{1}{c}{} &
\multicolumn{4}{c}{GOODNESS OF FIT} &
\multicolumn{1}{c}{} &
\multicolumn{3}{c}{IMPROVEMENT} \\
&&&&&&&&& \\                                                                         
\multicolumn{1}{c}{Cluster} &
\multicolumn{1}{c}{} &
\multicolumn{1}{c}{Model A} &
\multicolumn{1}{c}{Model B} &
\multicolumn{1}{c}{Model C} &
\multicolumn{1}{c}{Model D} &
\multicolumn{1}{c}{} &
\multicolumn{1}{c}{A$\rightarrow$ B } &
\multicolumn{1}{c}{B$\rightarrow$ CF } &
\multicolumn{1}{c}{B$\rightarrow$ 2T} \\
\hline                                                                               
&&&&&&&&& \\                                                                         
Abell 2744      & ~~~ & $9.9 \times 10^{-4}$ & 0.46                 & ---                     & ---                   & ~~~ &  1.00   & ---  & 0.98   \\       
Abell 478       & ~~~ & $8.6 \times 10^{-3}$ & $8.8 \times 10^{-3}$ & $5.2 \times 10^{-2}$    & 0.23         	      & ~~~ &  0.77   & 1.00 & 1.00  \\        
Abell 520       & ~~~ & 0.47                 & 0.48                 & ---                     & ---   	     	      & ~~~ &  0.69   & ---  & 0.57 \\    
Abell 586       & ~~~ & 0.87                 & 0.87                 & 0.88		      & 0.90  	     	      & ~~~ &  0.34   & 0.74 & 0.90  \\      
PKS0745-191     & ~~~ & $1.4 \times 10^{-3}$ & $5.4 \times 10^{-2}$ & 0.10                    & 0.22  	     	      & ~~~ &  1.00   & 1.00 & 1.00 \\      
Abell 665       & ~~~ & 0.43                 & 0.53                 & ---                     & ---   	     	      & ~~~ &  1.00   & ---  & 0.96  \\      
IRAS 09104+4109 & ~~~ & $7.0 \times 10^{-2}$ & $7.4 \times 10^{-2}$ & 0.11                    & 0.18  	     	      & ~~~ &  0.83   & 0.96 & 1.00  \\       
Abell 773       & ~~~ & $9.1 \times 10^{-3}$ & $1.8 \times 10^{-2}$ & ---                     & ---   	     	      & ~~~ &  1.00   & ---  & 0.82  \\    
Abell 963       & ~~~ & 0.24                 & 0.24                 & 0.23                    & 0.22  	     	      & ~~~ &  0.24   & 0.00 & 0.00 \\       
Zwicky 3146     & ~~~ & 0.61                 & 0.67                 & 0.81                    & 0.83  	     	      & ~~~ &  0.99   & 1.00 & 1.00   \\      
Abell 1068      & ~~~ & $6.0 \times 10^{-7}$ & $3.3 \times 10^{-2}$ & $4.5 \times 10^{-2}$    & 0.14  	              & ~~~ &  1.00   & 0.93 & 1.00 \\       
Abell 1413      & ~~~ & 0.19                 & 0.62                 & 0.65                    & 0.72  	     	      & ~~~ &  1.00   & 0.91 & 1.00  \\  
Abell 1689      & ~~~ & 0.80                 & 0.84                 & 0.85                    & 0.89  	     	      & ~~~ &  1.00   & 0.84 & 0.99  \\  
Abell 1704      & ~~~ & 0.92                 & 0.95                 & 0.96                    & 0.97  	     	      & ~~~ &  0.90   & 0.63 & 0.97   \\   
RXJ1347.5-1145  & ~~~ & $5.7 \times 10^{-2}$ & 0.20                 & 0.25                    & 0.26  	     	      & ~~~ &  1.00   & 0.98 & 0.96   \\      
Abell 1795      & ~~~ & $4.0 \times 10^{-5}$ & $4.3 \times 10^{-5}$ & $1.5 \times 10^{-4}$    & $1.3 \times 10^{-3}$  & ~~~ &  0.83   & 1.00 & 1.00   \\           
MS1358.4+6245   & ~~~ & 0.31                 & 0.67                 & 0.61                    & 0.69                  & ~~~ &  1.00   & 0.00 & 0.64 \\  
Abell 1835      & ~~~ & 0.64                 & 0.86                 & 0.89                    & 0.91                  & ~~~ &  1.00   & 0.99 & 1.00   \\       
MS1455.0+2232   & ~~~ & 0.56                 & 0.75                 & 0.75                    & 0.77                  & ~~~ &  1.00   & 0.52 & 0.80   \\       
Abell 2029      & ~~~ & $< 10^{-7}$          & $3.0 \times 10^{-6}$ & $4.7 \times 10^{-5}$    & $4.2 \times 10^{-4}$  & ~~~ &  1.00   & 1.00 & 1.00  \\      
Abell 2142      & ~~~ & $8.9 \times 10^{-3}$ & $7.3 \times 10^{-2}$ & $8.7 \times 10^{-2}$    & 0.15                  & ~~~ &  1.00   & 0.92 & 1.00   \\      
Abell 2163      & ~~~ & $6.7 \times 10^{-2}$ & 0.66                 & ---                     & ---                   & ~~~ &  1.00   & ---  & 0.98    \\       
Abell 2204      & ~~~ & $4.3 \times 10^{-5}$ & $1.7 \times 10^{-2}$ & $2.2 \times 10^{-2}$    & $2.9 \times 10^{-2}$  & ~~~ &  1.00   & 0.97 & 0.99    \\     
Abell 2218      & ~~~ & 0.33                 & 0.33                 & ---                     & ---                   & ~~~ &  0.55   & ---  & 0.50  \\       
Abell 2219      & ~~~ & $1.6 \times 10^{-2}$ & 0.54                 & ---		      & ---                   & ~~~ &  1.00   & ---  & 1.00   \\      
Abell 2261      & ~~~ & $5.3 \times 10^{-2}$ & 0.79                 & 0.84                    & 0.85                  & ~~~ &  1.00   & 0.97 & 0.99    \\      
Abell 2319      & ~~~ & $6.9 \times 10^{-2}$ & 0.12                 & ---                     & ---     	      & ~~~ &  1.00   & ---  & 1.00     \\      
MS2137.3-2353   & ~~~ & 0.96                 & 0.99                 & 0.99	              & 0.99    	      & ~~~ &  1.00   & 0.00 & 0.65     \\      
Abell 2390      & ~~~ & 0.73                 & 0.79                 & 0.79                    & 0.81   	              & ~~~ &  0.97   & 0.55 & 0.88    \\      
AC114           & ~~~ & $3.5 \times 10^{-2}$ & $9.3 \times 10^{-2}$ & ---		      & ---    		      & ~~~ &  1.00   & ---  & 0.20    \\      
&&&&&&&&& \\                                                                         
\hline 
&&&&&&&&& \\                                                                         
\end{tabular}
\end{center}
\parbox {7in}
{The goodness of fit results for spectral models A--D, listing 
the probabilities of exceeding the best-fitting reduced
$\chi^2$ values, assuming in each case that the model correctly describes 
the data. Column 6 lists the significance of the improvements to the fits  
obtained with model B over model A (\ie including the 
Galactic column density as a free parameter in the isothermal model).
Columns 7 and 8 list the significance of the improvements obtained by 
using the multiphase 
spectral models [either the cooling flow (CF) or
two-temperature models (2T)] in preference to the single-phase model B. 
}
\end{table*}

\clearpage

\begin{table*}
\vskip 0.2truein
\begin{center}
\caption{Results from the deprojection analysis of the ROSAT images}
\vskip 0.2truein
\begin{tabular}{ c c c c c c c c c c}
\hline                                                                                                                               
\multicolumn{1}{c}{} &
\multicolumn{1}{c}{} &
\multicolumn{1}{c}{} &
\multicolumn{2}{c}{MASS PROFILES} &
\multicolumn{1}{c}{} &
\multicolumn{4}{c}{DEPROJECTION RESULTS} \\
&&&&&&&&& \\                                                                                                                                                                   
Cluster         &   binsize       & ~ &  $\sigma$      & $r_{\rm c}$ & ~ &   $t_{\rm cool}$   &  $\overline{t_{\rm 100}}$ &   $r_{\rm cool}$  &   ${\dot M_{\rm I}}$             \\
                &   (arcsec/kpc)  & ~ &  (\kmps)        & (kpc)      & ~ &  ($10^{9}$ yr)   &     ($10^{9}$ yr)    &     (kpc)         &   (\Msunpyr) \\  
\hline                                                                                                                                                         
&&&&&&&&& \\                                                                                                                                                                         
Abell 2744      &  16/89.4        & ~ &  $1120^{+40}_{-50}$   & 450  & ~ &   $18.8^{+42.0}_{-6.4}$  & 19.2  & $<134$                   & $<101$ \\  
Abell 478       &  12/26.5        & ~ &  $860^{+70}_{-50}$    & 70   & ~ &   $1.1^{+0.1}_{-0.1}$    & 3.11  & $222^{+30}_{-50}$        & $810^{+176}_{-195}$   \\  
Abell 520       &  24/102         & ~ &  $980^{+70}_{-60}$    & 450  & ~ &   $16.7^{+24.0}_{-6.2}$  & 16.3  & $<154$                   & $<85$  \\  
Abell 586       &  16/60.3        & ~ &  $1050^{+450}_{-250}$ & 130  & ~ &   $5.5^{+0.7}_{-0.7}$    & 7.94  & $119^{+31}_{-29}$        & $159^{+69}_{-56}$     \\  
PKS0745-191     &  12/30.3        & ~ &  $930^{+90}_{-70}$    & 37.5 & ~ &   $1.1^{+0.1}_{-0.1}$    & 2.65  &   $178^{+19}_{-12}$      & $650^{+120}_{-148}$   \\  
Abell 665       &  30/118         & ~ &  $930^{+30}_{-30}$    & 200  & ~ &   $12.3^{+0.9}_{-0.9}$   & 13.0  & $<177$                   & $<232$ \\  
IRAS 09104+4109 &  12/81.0        & ~ &  $1120^{+330}_{-120}$ & 45   & ~ &   $2.0^{+0.1}_{-0.1}$    & 2.32  &   $193^{+90}_{-72}$      & $620^{+147}_{-87}$    \\  
Abell 773       &  16/71.5        & ~ &  $990^{+70}_{-50}$    & 180  & ~ &   $9.8^{+13.7}_{-6.6}$   & 11.2  & $<179$                   & $<238$ \\  
Abell 963       &  16/69.0        & ~ &  $750^{+50}_{-50}$    & 80   & ~ &   $4.1^{+1.3}_{-0.5}$    & 5.60  & $188^{+53}_{-84}$        & $340^{+129}_{-183}$   \\  
Zwicky 3146     &  12/64.8        & ~ &  $1150^{+280}_{-140}$ & 50   & ~ &   $1.7^{+0.1}_{-0.1}$    & 2.53  & $205^{+21}_{-43}$        & $1057^{+117}_{-146}$  \\  
Abell 1068      &  12/38.6        & ~ &  $780^{+90}_{-70}$    & 30   & ~ &   $1.2^{+0.1}_{-0.1}$    & 2.75  & $185^{+65}_{-50}$        & $408^{+128}_{-90}$    \\  
Abell 1413      &  30/98.7        & ~ &  $960^{+70}_{-50}$    & 130  & ~ &   $8.6^{+0.6}_{-0.7}$    & 8.63  & $119^{+29}_{-70}$        & $137^{+43}_{-85}$     \\  
Abell 1689      &  12/47.7        & ~ &  $990^{+60}_{-50}$    & 80   & ~ &   $2.9^{+0.5}_{-0.2}$    & 4.34  & $180^{+35}_{-61}$        & $563^{+185}_{-174}$   \\  
Abell 1704      &  12/53.5        & ~ &  $750^{+220}_{-90}$   & 30   & ~ &   $2.1^{+0.1}_{-0.2}$    & 3.59  & $187^{+53}_{-54}$        & $306^{+92}_{-73}$     \\  
RXJ1347.5-1145  &  12/81.7        & ~ &  $1850^{+270}_{-500}$ & 75   & ~ &   $2.6^{+0.2}_{-0.1}$    & 3.00  & $196^{+90}_{-73}$        & $1378^{+75}_{-316}$   \\  
Abell 1795      &  16/26.3        & ~ &  $740^{+20}_{-20}$    & 50   & ~ &   $1.4^{+0.1}_{-0.1}$    & 3.42  & $188^{+11}_{-16}$        & $449^{+46}_{-46}$     \\  
MS1358.4+6245   &   8/46.3        & ~ &  $830^{+340}_{-80}$   & 40   & ~ &   $2.8^{+1.2}_{-0.6}$    & 4.76  & $113^{+187}_{-44}$       & $123^{+369}_{-50}$    \\  
Abell 1835      &   8/39.5        & ~ &  $1000^{+120}_{-70}$  & 50   & ~ &   $1.5^{+0.5}_{-0.3}$    & 2.44  & $231^{+26}_{-14}$        & $1154^{+432}_{-482}$  \\  
MS1455.0+2232   &   8/40.1        & ~ &  $760^{+190}_{-70}$   & 45   & ~ &   $1.1^{+0.2}_{-0.1}$    & 2.15  & $215^{+45}_{-35}$        & $712^{+185}_{-53}$    \\  
Abell 2029      &  16/31.5        & ~ &  $880^{+40}_{-30}$    & 80   & ~ &   $1.5^{+0.1}_{-0.1}$    & 3.54  & $193^{+42}_{-20}$        & $576^{+99}_{-79}$     \\  
Abell 2142      &  16/35.7        & ~ &  $910^{+70}_{-40}$    & 100  & ~ &   $4.3^{+0.8}_{-0.7}$    & 5.38  & $154^{+42}_{-29}$        & $303^{+167}_{-72}$    \\   
Abell 2163      &  12/52.1        & ~ &  $1210^{+30}_{-40}$   & 300  & ~ &   $12.7^{+12.3}_{-4.6}$  & 12.1  & $<130$                   & $<90$ \\
Abell 2204      &  12/41.4        & ~ &  $1020^{+130}_{-60}$  & 25   & ~ &   $0.94^{+0.04}_{-0.04}$ & 2.01  & $235^{+34}_{-90}$        & $1007^{+98}_{-263}$   \\  
Abell 2218      &  16/61.4        & ~ &  $830^{+30}_{-40}$    & 230  & ~ &   $10.4^{+4.1}_{-2.0}$   & 12.8  & $<153$                   & $<133$                \\  
Abell 2219      &  10/46.2        & ~ &  $1050^{+50}_{-50}$   & 250  & ~ &   $9.0^{+18.8}_{-3.8}$   & 10.7  & $<254$                   & $<485$ \\  
Abell 2261      &  8/36.6         & ~ &  $1030^{+240}_{-120}$ & 80   & ~ &   $3.0^{+1.4}_{-0.6}$    & 5.63  & $125^{+76}_{-34}$        & $227^{+238}_{-87}$    \\  
Abell 2319      &  32/47.4        & ~ &  $890^{+30}_{-20}$    & 150  & ~ &   $9.5^{+10.4}_{-3.2}$   & 9.81  & $109^{+55}_{-39}$        & $80^{+81}_{-47}$      \\  
MS2137.3-2353   &  12/67.7        & ~ &  $830^{+140}_{-50}$   & 32   & ~ &   $1.2^{+0.1}_{-0.1}$    & 1.65  & $216^{+156}_{-47}$       & $754^{+263}_{-117}$   \\  
Abell 2390      &  16/75.1        & ~ &  $1190^{+520}_{-240}$ & 60   & ~ &   $4.2^{+0.3}_{-0.2}$    & 5.45  & $146^{+41}_{-34}$        & $370^{+97}_{-69}$     \\  
AC114           &  12/67.6        & ~ &  $1030^{+60}_{-40}$   & 300  & ~ &   $16.9^{+32.7}_{-6.2}$  & 16.6  & $<135$                   & $<120$ \\  
&&&&&&&& \\                                                                                                                                                           
\hline                                                                                                                              
&&&&&&&& \\                                                                                                                                                           
\end{tabular}			    
\end{center}
\parbox {7in}
{Notes: Column 2 lists the binsize (in arcsec and kpc) used in the analysis. 
Columns 3 and 4 summarize the velocity dispersions and core
radii used to parameterize the mass profiles (Section 4). 
Columns 5 and 6 list the cooling times {\it i.e.} the time for the gas to 
cool from the ambient cluster temperature at constant pressure 
(in units of $10^9$ yr) within the central bin and within a fixed radius of 
100 kpc ($t_{\rm cool}$ and $\overline{t_{\rm 100}}$, respectively). Cooling 
radii ($r_{\rm cool}$) are the radii where the cooling time of the cluster gas 
becomes equal to the age of the Universe 
($1.3 \times 10^{10}$ yr). Mass deposition 
rates (${\dot M_{\rm I}}$) are the integrated mass deposition rates within 
the cooling radii in \Msunpyr. 
Errors on the velocity dispersions are 90 per cent confidence limits. 
Errors on the cooling times are the 10 and 90 percentile values from 100 
Monte Carlo simulations, using the best-fit mass models for the clusters. The 
upper and lower confidence limits on the cooling radii are the
points where the 10 and 90 percentiles exceed and become less than the Hubble
time, respectively. Errors on the mass deposition rates are the 90 and 10
percentile values at the upper and lower limits for the cooling radius.
For Abell 478, 586, 1795, 2029 and 2219, the introduction of a second
mass component was found to improve the isothermality of the 
temperature profile and for these clusters the multi-component mass models 
(described in Section 4) have been used in determining the deprojection 
results. We note that the errors on the deprojection results do not account 
for the uncertainties in the total mass profiles.}
\end{table*}

\clearpage

\begin{table*}
\vskip 0.2truein
\begin{center}
\caption{Absorption-corrected mass deposition rates }
\vskip 0.2truein
\begin{tabular}{ c c c c c c }
\hline                                                                                                                               
\multicolumn{1}{c}{} &
\multicolumn{1}{c}{} &
\multicolumn{1}{c}{INTRINSIC} &
\multicolumn{1}{c}{ORIGINAL} &
\multicolumn{1}{c}{CORRECTED} &
\multicolumn{1}{c}{SPECTRAL} \\
\multicolumn{1}{c}{} &
\multicolumn{1}{c}{} &
\multicolumn{1}{c}{ABSORPTION} &
\multicolumn{1}{c}{DEPROJ.} &
\multicolumn{1}{c}{DEPROJ.} &
\multicolumn{1}{c}{ANALYSIS} \\
&&&& \\       
Cluster         & ~ & $\Delta N_{\rm H}$       &  ${\dot M_{\rm I}}$ & ${\dot M_{\rm C}}$ & ${\dot M_{\rm S}}$         \\
                & ~ & ($10^{21}$ \apc)         &  (\Msunpyr) & (\Msunpyr)  & (\Msunpyr)   \\
\hline                                                                                                                    
&&&& \\       
Abell 478       & ~ &  $1.7^{+0.2}_{-0.3}$    &  $810^{+176}_{-195}$   & $1059^{+421}_{-300}$   &  $1347^{+267}_{-317}$        \\   
PKS0745-191     & ~ &  $2.8^{+1.1}_{-1.3}$    &  $650^{+120}_{-148}$   & $950^{+166}_{-200}$   &  $1455^{+356}_{-510}$   	  \\  
IRAS 09104+4109 & ~ &  $3.0^{+1.6}_{-1.3}$    &  $620^{+147}_{-87}$    & $1060^{+356}_{-222}$   &  $1655^{+645}_{-718}$  	  \\  
Zwicky 3146     & ~ &  $1.2^{+0.3}_{-0.3}$    &  $1057^{+117}_{-146}$  & $1358^{+449}_{-205}$   &  $2228^{+357}_{-636}$  	  \\  
Abell 1068      & ~ &  $3.2^{+1.0}_{-0.8}$    &  $408^{+128}_{-90}$    & $937^{+185}_{-186}$    &  $1345^{+145}_{-177}$  	  \\  
*Abell 1413     & ~ &  $3.4^{+1.3}_{-0.8}$    &  $137^{+43}_{-85}$     & $321^{+173}_{-55}$    &  $644^{+156}_{-159}$  	  \\  
Abell 1689      & ~ &  $4.1^{+5.6}_{-1.8}$    &  $563^{+185}_{-174}$   & $2016^{+1575}_{-908}$  &  $351^{+291}_{-209}$  	  \\ 
Abell 1704      & ~ &  $3.3^{+5.2}_{-1.6}$    &  $306^{+92}_{-73}$     & $632^{+793}_{-158}$   &  $987^{+382}_{-616}$  	  \\  
RXJ1347.5-1145  & ~ &  $2.7^{+3.0}_{-0.7}$    &  $1378^{+75}_{-316}$   & $1790^{+413}_{-411}$   &  $3479^{+335}_{-1151}$  	  \\  
Abell 1795      & ~ &  $2.1^{+0.6}_{-0.4}$    &  $449^{+46}_{-46}$     & $1038^{+229}_{-144}$    &  $300^{+104}_{-110}$  	  \\  
MS1358.4+6245   & ~ &  $6.4^{+8.7}_{-3.8}$    &  $123^{+369}_{-50}$    & $703^{+1668}_{-445}$   &  $691^{+348}_{-287}$  	  \\  
Abell 1835      & ~ &  $3.2^{+1.6}_{-0.8}$    &  $1154^{+432}_{-482}$  & $2111^{+2154}_{-921}$ &  $1761^{+523}_{-593}$  	  \\  
MS1455.0+2232   & ~ &  $4.1^{+4.3}_{-1.6}$    &  $712^{+185}_{-53}$    & $1227^{+715}_{-172}$  &  $1728^{+714}_{-848}$  	  \\  
Abell 2029      & ~ &  $3.1^{+0.5}_{-0.4}$    &  $576^{+99}_{-79}$     & $1220^{+223}_{-160}$   & $547^{+82}_{-81}$  	  \\  
Abell 2142      & ~ &  $3.6^{+1.8}_{-1.0}$    &  $303^{+167}_{-72}$    & $1303^{+409}_{-653}$    & $529^{+177}_{-164}$  	  \\  
Abell 2204      & ~ &  $3.3^{+0.9}_{-0.8}$    &  $1007^{+98}_{-263}$   & $1660^{+634}_{-232}$   &  $2103^{+356}_{-378}$  	  \\  
Abell 2261      & ~ &  $3.8^{+1.8}_{-1.4}$    &  $227^{+238}_{-87}$    & $680^{+248}_{-164}$    & $1323^{+215}_{-263}$   	  \\  
MS2137.3-2353   & ~ &  $6.2^{+6.1}_{-2.8}$    &  $754^{+263}_{-117}$   & $1537^{+1499}_{-399}$  &  $1467^{+880}_{-726}$  	  \\  
Abell 2390      & ~ &  $2.9^{+7.6}_{-1.5}$    &  $370^{+97}_{-69}$     & $599^{+852}_{-163}$   &  $1533^{+581}_{-1112}$  	  \\  

&&&& \\       
\hline                                                                                                                                                   
&&&& \\       
\end{tabular}
\end{center}
\parbox {7in}
{A comparison of the mass deposition rates measured  from the
deprojection analysis, corrected for the effects of intrinsic absorption
due to cold gas (${\dot M_{\rm C}}$), with 
the values determined from the spectral
analysis (${\dot M_{\rm S}}$). Errors on the corrected deprojection values 
account for the statistical uncertainties in both the 
deprojected quantities and intrinsic column densities. 
No entries are listed for Abell 586 or 963 since the spectral 
data do not constrain the mass deposition rates and intrinsic 
column densities for these systems. }
\end{table*}

\clearpage

\begin{table}
\vskip 0.2truein
\begin{center}
\caption{Cooling flow flux fractions}
\vskip 0.2truein
\begin{tabular}{ c c c }
\hline                                             
Cluster         & ~ & ($L_{\rm cool}/L_{\rm X}$)  \\
\hline       
&& \\       
Abell 478       & ~ & $0.37^{+0.15}_{-0.12}$ \\  
PKS0745-191     & ~ & $0.36^{+0.19}_{-0.17}$ \\  
IRAS 09104+4109 & ~ & $0.55^{+0.45}_{-0.32}$ \\  
Zwicky 3146     & ~ & $0.63^{+0.37}_{-0.31}$ \\  
Abell 1068      & ~ & $0.74^{+0.26}_{-0.20}$ \\  
Abell 1413      & ~ & $0.30^{+0.13}_{-0.09}$ \\  
Abell 1689      & ~ & $0.10^{+0.10}_{-0.06}$ \\ 
Abell 1704      & ~ & $0.48^{+0.52}_{-0.36}$ \\  
RXJ1347.5-1145  & ~ & $0.90^{+0.10}_{-0.57}$ \\  
Abell 1795      & ~ & $0.11^{+0.05}_{-0.05}$ \\  
MS1358.4+6245   & ~ & $0.38^{+0.62}_{-0.18}$ \\  
Abell 1835      & ~ & $0.34^{+0.22}_{-0.15}$ \\  
MS1455.0+2232   & ~ & $0.40^{+0.60}_{-0.24}$ \\  
Abell 2029      & ~ & $0.21^{+0.04}_{-0.04}$ \\  
Abell 2142      & ~ & $0.13^{+0.07}_{-0.05}$ \\  
Abell 2204      & ~ & $0.49^{+0.25}_{-0.14}$ \\  
Abell 2261      & ~ & $0.59^{+0.18}_{-0.20}$ \\  
MS2137.3-2353   & ~ & $0.29^{+0.40}_{-0.16}$ \\  
Abell 2390      & ~ & $0.51^{+0.49}_{-0.42}$ \\  
&& \\       
\hline                                                                                                                                                   
&& \\       
\end{tabular}
\end{center}
\parbox {3.5in}
{Notes: The fractions of the $2-10$ keV X-ray luminosities contributed 
by the cooling flows (determined with spectral model C and measured 
in the G3 detector). The errors are 90 per cent confidence limits.} 
\end{table}

\clearpage

\begin{table*}
\vskip 0.2truein
\begin{center}
\caption{The column densities of intrinsic X-ray absorbing material} 
\vskip 0.2truein
\begin{tabular}{ c c c c c c c c c }
\hline                                                               
\multicolumn{1}{c}{} &
\multicolumn{5}{c}{CF CLUSTERS} &
\multicolumn{1}{c}{} &
\multicolumn{2}{c}{NCFs} \\
&&&&&&&& \\							                                                      
~~~~~
Cluster           &        B-A               &     C-A              &          C'-A          &  D-A                   & $f$          &~~~~~~ &  Cluster               &      B-A       \\
\hline                                                                                                                	   
&&&&&&&& \\							                                                      	   
Abell 478         & $-0.06^{+0.08}_{-0.09}$  & $1.7^{+0.2}_{-0.3}$  & $0.83^{+0.28}_{-0.38}$ & $9.4^{+1.6}_{-1.4}$    & $>0.85$    & ~~~~~~ &  Abell 2744   & $1.28^{+0.21}_{-0.21}$ \\
Abell 586         & $0.07^{+0.35}_{-0.32}$   &      ---             & $0.85^{+0.54}_{-0.93}$ & $12.3^{+67.8}_{-11.1}$ &  U.C.      & ~~~~~~ &  Abell 520    & $0.14^{+0.23}_{-0.21}$ \\ 
PKS0745-191$^*$   & $0.19^{+0.10}_{-0.10}$   & $2.8^{+1.1}_{-1.3}$  & $1.09^{+0.29}_{-0.57}$ & $8.0^{+15.0}_{-3.5}$   & $>0.89$    & ~~~~~~ &  Abell 665    & $0.30^{+0.14}_{-0.14}$ \\ 
IRAS 09104+4109   & $0.21^{+0.26}_{-0.24}$   & $3.0^{+1.6}_{-1.3}$  & $0.93^{+0.36}_{-0.80}$ & $8.9^{+63.1}_{-5.6}$   & $>0.79$    & ~~~~~~ &  Abell 773    & $0.28^{+0.14}_{-0.14}$ \\ 
Abell 963         & $0.02^{+0.14}_{-0.15}$   &       ---            & $0.01^{+0.21}_{-0.11}$ &       ---              &  U.C.      & ~~~~~~ &  Abell 2163   & $0.66^{+0.13}_{-0.12}$ \\ 
Zwicky 3146       & $-0.19^{+0.11}_{-0.10}$  & $1.2^{+0.3}_{-0.3}$  & $0.53^{+0.17}_{-0.26}$ & $0.4^{+1.6}_{-0.4}$    & $>0.42$    & ~~~~~~ &  Abell 2218   & $-0.06^{+0.14}_{-0.14}$ \\
Abell 1068        & $0.85^{+0.15}_{-0.14}$   & $3.2^{+1.0}_{-0.8}$  & $1.19^{+0.44}_{-0.39}$ & $1.7^{+1.1}_{-0.4}$    & $>0.95$    & ~~~~~~ &  Abell 2219   & $0.59^{+0.10}_{-0.10}$ \\ 
Abell 1413        & $0.44^{+0.10}_{-0.11}$   & $3.4^{+1.3}_{-0.8}$  & $0.53^{+0.36}_{-0.19}$ & $3.9^{+3.6}_{-2.5}$    & $>0.89$    & ~~~~~~ &  Abell 2319   & $0.25^{+0.11}_{-0.11}$ \\ 
Abell 1689        & $0.14^{+0.09}_{-0.08}$   & $4.1^{+5.6}_{-1.8}$  & $0.19^{+0.12}_{-0.24}$ & $16.2^{+12.6}_{-5.5}$  & $>0.59$    & ~~~~~~ &  AC114        & $0.51^{+0.22}_{-0.21}$ \\ 
Abell 1704        & $0.46^{+0.31}_{-0.29}$   & $3.3^{+5.2}_{-1.6}$  & $0.45^{+1.38}_{-0.24}$ & $10.9^{+12.1}_{-9.4}$  & $>0.79$    & ~~~~~~ &               &                        \\
RXJ1347.5-1145    & $0.49^{+0.16}_{-0.15}$   & $2.7^{+3.0}_{-0.7}$  & $1.05^{+0.17}_{-0.17}$ & $8.3^{+13.2}_{-5.3}$   & $>0.58$    & ~~~~~~ &               &                        \\
Abell 1795        & $0.05^{+0.05}_{-0.05}$   & $2.1^{+0.6}_{-0.4}$  & $0.17^{+0.11}_{-0.11}$ & $0.7^{+1.0}_{-0.5}$    & $>0.76$    & ~~~~~~ &               &                        \\
MS1358.4+6245     & $0.85^{+0.32}_{-0.29}$   & $6.4^{+8.7}_{-3.8}$  & $0.97^{+0.55}_{-0.22}$ & $1.8^{+1.9}_{-0.7}$    & $>0.80$    & ~~~~~~ &               &                        \\
Abell 1835        & $0.32^{+0.09}_{-0.09}$   & $3.2^{+1.6}_{-0.8}$  & $0.63^{+0.34}_{-0.35}$ & $3.1^{+3.1}_{-1.6}$    & $>0.80$    & ~~~~~~ &               &                        \\
MS1455.0+2232     & $0.41^{+0.17}_{-0.17}$   & $4.1^{+4.3}_{-1.6}$  & $0.65^{+0.64}_{-0.36}$ & $1.8^{+7.5}_{-1.1}$    & $>0.63$    & ~~~~~~ &               &                        \\
Abell 2029        & $0.27^{+0.04}_{-0.05}$   & $3.1^{+0.5}_{-0.4}$  & $0.45^{+0.10}_{-0.09}$ & $6.6^{+2.1}_{-2.1}$    & $>0.91$    & ~~~~~~ &               &                        \\
Abell 2142        & $0.35^{+0.10}_{-0.10}$   & $3.6^{+1.8}_{-1.0}$  & $0.35^{+0.27}_{-0.08}$ & $15.0^{+7.5}_{-11.4}$  & $>0.80$    & ~~~~~~ &               &                        \\
Abell 2204        & $0.61^{+0.12}_{-0.12}$   & $3.3^{+0.9}_{-0.8}$  & $1.02^{+0.37}_{-0.38}$ & $1.9^{+1.4}_{-0.7}$    & $>0.93$    & ~~~~~~ &               &                        \\
Abell 2261        & $0.84^{+0.17}_{-0.17}$   & $3.8^{+1.8}_{-1.4}$  & $1.51^{+0.39}_{-0.61}$ & $5.9^{+3.5}_{-2.7}$    & $>0.84$    & ~~~~~~ &               &                        \\
MS2137.3-2353     & $0.59^{+0.29}_{-0.28}$   & $6.2^{+6.1}_{-2.8}$  & $0.60^{+0.39}_{-0.19}$ & $3.3^{+47.7}_{-2.5}$   & $>0.87$    & ~~~~~~ &               &                        \\
Abell 2390        & $0.43^{+0.32}_{-0.30}$   & $2.9^{+7.6}_{-1.5}$  & $1.17^{+0.42}_{-0.88}$ & $5.5^{+14.9}_{-4.0}$   & $>0.57$    & ~~~~~~ &               &                        \\
&&&&&&&& \\							                                                      	   
\hline                                                                                                                                                                                                      
MEAN              & $0.35 \pm 0.30$          & $3.4 \pm 1.3$        &  $0.72 \pm 0.39$       &   $6.3 \pm 4.7$        &  & ~ &     MEAN      &  $0.44 \pm 0.39$         \\
\hline                                                                                                                                                   
&&&&&&&& \\
\end{tabular}
\end{center}
\parbox {7in}
{A summary of the results on intrinsic absorption 
from the ASCA data. Columns $2-5$ list the differences between the 
spectrally-determined column densities and the nominal Galactic values 
for spectral models B, C, D and C'. Also listed are the (90 per cent
confidence) constraints on the covering fraction of the intrinsic
absorber, $f$,  using spectral model C. All results assume solar metallicity in
the absorbing material. For PKS0745-191$^*$, the ASCA data indicate a Galactic 
column density $3.5 \times 10^{21}$ \apc~which is lower than the 
nominal value of $4.24 \times 10^{21}$ \apc~(Dickey
\& Lockman 1990). The lower value of $3.5 \times 10^{21}$ \apc~has been 
used in the calculation of the intrinsic column densities. The errors on the 
mean values are the standard deviations of the distributions.
}
\end{table*}

\clearpage

\begin{table*}
\vskip 0.2truein
\begin{center}
\caption{The masses of intrinsic X-ray absorbing gas} 
\vskip 0.2truein
\begin{tabular}{ c c c c c c }
\hline                                                               
\multicolumn{1}{c}{} &
\multicolumn{1}{c}{$r_{\rm abs}$} &
\multicolumn{1}{c}{$M_{\rm abs}$ [$Z\sun$]} &
\multicolumn{1}{c}{$M_{\rm abs}$ [$Z_{\rm ICM}$]} &
\multicolumn{1}{c}{${\dot M(r<r_{\rm abs})}$} &
\multicolumn{1}{c}{$M_{\rm acc}$} \\
Cluster           &    (kpc)          &  $(10^{11} M_{\sun})$           & $(10^{11} M_{\sun})$      & ($10^9$ \Msunpyr)         &       $(10^{11} M_{\sun})$      \\
\hline                                                                                                               
&&&&& \\
Abell 478         & $115^{+31}_{-22}$ & $ 7.30^{+ 4.60}_{ -2.61}$ &  $ 14.60^{+  9.21}_{ -5.22}$   & $628^{+204}_{-150}$     &  $15.70^{+ 5.10}_{ -3.75}$ \\
PKS0745-191       & $121^{+45}_{-45}$ & $13.31^{+12.72}_{ -8.30}$ &  $ 26.61^{+ 25.44}_{-16.60}$   & $909^{+144}_{-344}$     &  $22.73^{+ 3.60}_{ -8.60}$ \\ 
IRAS 09104+4109   & $120^{+82}_{-80}$ & $14.02^{+27.83}_{-12.53}$ &  $ 28.05^{+ 55.66}_{-25.06}$   & $911^{+292}_{-361}$     &  $22.77^{+ 7.30}_{ -9.03}$ \\ 
Zwicky 3146       & $119^{+42}_{-22}$ & $ 5.52^{+ 4.83}_{ -1.94}$ &  $ 11.03^{+  9.67}_{ -3.89}$   & $1050^{+269}_{-145}$    &  $26.25^{+ 6.73}_{ -3.63}$ \\
Abell 1068        & $135^{+38}_{-39}$ & $18.93^{+13.13}_{ -9.60}$ &  $ 37.86^{+ 26.26}_{-19.19}$   & $769^{+232}_{-223}$     &  $19.23^{+ 5.80}_{ -5.58}$ \\ 
Abell 1689        & $123^{+44}_{-43}$ & $20.13^{+22.05}_{-11.99}$ &  $ 40.27^{+ 44.10}_{-23.98}$   & $862^{+890}_{-347}$     &  $21.55^{+22.25}_{ -8.68}$ \\ 
Abell 1704        & $99^{+69}_{-19}$  & $10.50^{+24.50}_{ -3.98}$ &  $ 21.00^{+ 49.00}_{ -7.95}$   & $386^{+536}_{-106}$     &  $ 9.65^{+13.40}_{ -2.65}$ \\
RXJ1347.5-1145    & $114^{+90}_{-73}$ & $11.39^{+29.14}_{ -9.95}$ &  $ 22.78^{+ 58.27}_{-19.91}$   & $1440^{+708}_{-748}$    &  $36.00^{+17.70}_{-18.70}$ \\
Abell 1795        & $124^{+22}_{-29}$ & $10.48^{+ 4.46}_{ -4.45}$ &  $ 20.96^{+  8.93}_{ -8.89}$   & $645^{+137}_{-163}$     &  $16.13^{+ 3.43}_{ -4.08}$ \\
MS1358.4+6245     & $79^{+65}_{-50}$  & $12.97^{+35.97}_{-11.32}$ &  $ 25.93^{+ 71.94}_{-22.64}$   & $232^{+811}_{-133}$     &  $ 5.80^{+20.27}_{ -3.33}$ \\
Abell 1835        & $129^{+49}_{-67}$ & $17.29^{+17.27}_{-13.39}$ &  $ 34.57^{+ 34.54}_{-26.78}$   & $1430^{+1307}_{-632}$   &  $35.75^{+32.67}_{-15.80}$ \\
MS1455.0+2232     & $181^{+39}_{-81}$ & $43.60^{+27.57}_{-30.81}$ &  $ 87.20^{+ 55.14}_{-61.62}$   & $1596^{+749}_{-613}$    &  $39.90^{+18.73}_{-15.32}$ \\
Abell 2029        & $126^{+15}_{-17}$ & $15.98^{+ 4.35}_{ -4.17}$ &  $ 31.95^{+  8.71}_{ -8.35}$   & $839^{+152}_{-128}$     &  $20.98^{+ 3.80}_{ -3.20}$ \\
Abell 2142        & $93^{+32}_{-40}$  & $10.11^{+ 9.06}_{ -6.92}$ &  $ 20.21^{+ 18.13}_{-13.83}$   & $374^{+219}_{-201}$     &  $ 9.35^{+ 5.47}_{ -5.03}$ \\
Abell 2204        & $136^{+50}_{-33}$ & $19.81^{+18.26}_{ -8.72}$ &  $ 39.63^{+ 36.51}_{-17.45}$   & $1337^{+415}_{-271}$    &  $33.43^{+10.38}_{ -6.78}$ \\
Abell 2261        & $94^{+34}_{-76}$  & $10.90^{+10.27}_{-10.51}$ &  $ 21.80^{+ 20.54}_{-21.03}$   & $329^{+201}_{-301}$     &  $ 8.23^{+ 5.03}_{ -7.53}$ \\
MS2137.3-2353     & $170^{+67}_{-68}$ & $58.16^{+66.00}_{-38.17}$ &  $116.32^{+132.00}_{-76.34}$   & $1643^{+1162}_{-563}$   &  $41.08^{+29.05}_{-14.08}$ \\
Abell 2390        & $62^{+50}_{-25}$  & $ 3.62^{+11.28}_{ -2.40}$ &  $  7.24^{+ 22.57}_{ -4.79}$   & $293^{+573}_{-99}$      &  $ 7.33^{+14.33}_{ -2.47}$ \\
&&&&& \\
\hline                                                                                                                                                                                                   
&&&&& \\
\end{tabular}
\end{center}
\parbox {7in}
{Notes: $r_{\rm abs}$ values are the radii at which the
cooling time of the gas first exceeds $5 \times 10^9$ yr, determined from
the absorption-corrected deprojection results. $M_{\rm abs}$ 
values are the masses of absorbing material within radii $r_{\rm abs}$ 
implied by the observed intrinsic column densities (determined with spectral
model C) assuming the absorption to be due to cold gas with solar 
metallicity (column 3) or $Z=0.4Z_{\odot}$ (column 4). Column 5 lists the 
integrated mass deposition rates (corrected for the effects of absorption) 
within radii $r_{\rm abs}$. $M_{\rm acc}$ values are the masses of material
accumulated within radii $r_{\rm abs}$ during the first 5 Gyr. Errors 
incorporate the statistical uncertainties in both the intrinsic column 
densities (determined with spectral model C) and the deprojection results. 
Errors on the $r_{\rm abs}$ values are the 10 and 90 percentile values from 
100 Monte Carlo simulations. Errors on the  $M_{\rm abs}$ values are the 
extrema incorporating the 
joint errors in $r_{\rm abs}$ and $\Delta N_{\rm abs}$. Results are not 
tabulated for Abell 1413 since the central cooling time in this cluster 
exceeds 5 Gyr. }
\end{table*}

\clearpage

\begin{table*}
\vskip 0.2truein
\begin{center}
\caption{The reprocessed luminosities and observed IRAS fluxes} 
\vskip 0.2truein
\begin{tabular}{ c c c c c c }
\hline                                                               
\multicolumn{1}{c}{} &
\multicolumn{1}{c}{$L_{\rm repro.}$} &
\multicolumn{1}{c}{$S_{\rm 60}$} &
\multicolumn{1}{c}{$S_{\rm 100}$} &
\multicolumn{1}{c}{$\Delta R$} & 
\multicolumn{1}{c}{$L_{\rm 1-1000\mu m}$} \\
Cluster      &  ($10^{44}$ \ergps)  & (mJy) & (mJy) & (arcmin)   &  ($10^{44}$ \ergps)  \\
\hline                                             	      
&&&&& \\						      
Abell 478          & 10.5  & $570\pm58$ & $5490\pm440$ & 0.55  &  $60\pm19$   \\
PKS0745-191        & 13.4  & $<294$     & $<1680$      &  ---  &  $<29.0$  \\
IRAS 09104+4109    & 13.3  & $560\pm39$ & $<324$       &  ---  &  $<390$   \\
Zwicky 3146        & 6.15  & $<111$     & $880\pm56$   & 1.53  &  $<111$ \\
Abell 1068         & 9.05  & $680\pm38$ & $950\pm108$  & 0.02  &  $59\pm18$   \\
Abell 1413         & 5.71  & $<111$     & $<333$       &  ---  &  $<14.2$  \\
Abell 1689         & 3.46  & $<141$     & $<357$       &  ---  &  $<27.3$  \\
Abell 1704         & 6.84  & $90\pm41$  & $280\pm83$   & 0.43  &  $27\pm11$   \\
RXJ1347.5-1145     & 49.7  & $<141$     & $<498$       &  ---  &  $<196$   \\
Abell 1795         & 1.95  & $<144$     & $<354$       &  ---  &  $<3.23$  \\
MS1358.4+6245      & 6.45  & $<81$      & $<222$       &  ---  &  $<51.6$  \\
Abell 1835         & 16.7  & $140\pm64$ & $<504$       & 0.05  &  $<61.5$  \\
MS1455.0+2232      & 12.3  & $<96$      & $<378$       &  ---  &  $<46.6$  \\
Abell 2029         & 5.13  & $<96$      & $<351$       &  ---  &  $<3.98$  \\
Abell 2142         & 5.47  & $130\pm51$ & $480\pm156$  & 2.50  &  $7.2\pm2.9$   \\
Abell 2204         & 19.2  & $<288$     & $1760\pm306$ & 2.58  &  $<64.8$  \\
Abell 2261         & 14.9  & $150\pm30$ & $410\pm53$   & 1.66  &  $45\pm15$   \\
MS2137.3-2353      & 11.0  & $<69$      & $<249$       & ---   &  $<46.9$  \\
Abell 2390         & 17.4  & $130\pm60$ & $750\pm207$  & 1.73  &  $66\pm26$   \\

&&&&& \\	    					  
\hline              					  
&&&&& \\
\end{tabular}
\end{center}
\parbox {7in}
{Notes: Column 2 lists the X-ray luminosities expected to be reprocessed into 
far infrared emission by the intrinsic absorbing material in the 
clusters. The best-fit parameters determined with spectral model C have been 
used in the calculations. Columns 3 and 4 lists the observed $60$ and 
$100\mu$m IRAS fluxes within a 4 arcmin (radius) aperture 
centered on the X-ray centres (Table 3), determined with the IPAC SCANPI software and 
co-added IRAS scans (we use the median of the coadded scans). Error bars are 
the root-mean-square deviations in the residuals, external to the source 
extraction regions, after baseline subtraction.  
Where no clear detection was made, an upper limit equal to three times the r.m.s.
deviation in the residuals is given (although systematic errors could
exceed these limits in individual cases). Where a possible detection is
made, Column 6 lists the in-scan separation, $\Delta R$ 
(in arcmin), between the peak of the $100\mu$m emission (or $60\mu$m
emission in the case of Abell 1835) and the X-ray centre. 
Column 7 lists the total $1-1000\mu$m luminosities calculated using
equation 4.}
\end{table*}

\end{document}